\documentclass[a4paper,11pt]{article}
\pdfoutput=1 
\usepackage{jcappub} 

\usepackage[T1]{fontenc} 
\graphicspath{{figures/}}

\title{\boldmath Charge-dependent atmospheric muon flux at 17 GV geomagnetic cutoff with the mini-ICAL detector}

\author[a]{S. Pethuraj}
\author[a]{Raj Shah}
\author[a]{G. Majumder}
\author[b]{J. M. John}

\affiliation[a]{Tata Institute of Fundamental Research,\\Mumbai, India-400005}
\affiliation[b]{Institute of Nuclear Physics, Polish Academy of Sciences,\\Krakow, Poland}

\emailAdd{spethuraj135@gmail.com}
\emailAdd{rajbhupen20@gmail.com}
\emailAdd{gobinda@tifr.res.in}
\emailAdd{jimmjohn007@gmail.com }

\abstract{
The Iron CALorimeter (ICAL) detector at the India-Based Neutrino
Observatory (INO) was conceived as an underground experiment designed
to measure atmospheric neutrino oscillation parameters. As part of the
R\&D programme, a scaled prototype (mini-ICAL), 85\,ton, approximately
1/600$^{\mathrm{th}}$ the mass of the full detector, was
constructed at the IICHEP Transit Campus, Madurai (altitude 150\,m;
latitude 9.9372$^\circ$\,N; longitude 78.013$^\circ$\,E; geomagnetic
latitude 1.44$^\circ$\,N; vertical cutoff rigidity 17\,GV) and operated
between 2018 and 2022.
The prototype enabled measurements of charge-dependent cosmic muon
spectra in the vicinity of the geomagnetic equator and provided an
important validation of detector performance, reconstruction
algorithms, and simulation frameworks for the ICAL experiment.
Differential fluxes of
$\mu^{-}$ and $\mu^{+}$ were measured over the momentum range
$\sim$\,1--5\,GeV/c. The obtained momentum spectra are
systematically lower than those reported at sites with smaller
geomagnetic cutoff rigidities, consistent with the suppression of low-
and intermediate-rigidity primary cosmic rays at the 17\,GV
cutoff. The measurements are compared with predictions from
different hadronic interaction models available in CORSIKA simulations. 
}

\begin{document}
\maketitle
\flushbottom
\section{Introduction}
\label{sec:intro}

Cosmic rays originate from the Sun as well as from astrophysical
sources beyond the solar system and span energies from a few
$10^{7}$\,eV to beyond $10^{20}$\,eV at the Earth's atmosphere. The structure of their energy
spectrum encodes information on their origin and acceleration
mechanisms. The primary flux is dominated by protons ($\sim$90\%) and
helium nuclei ($\sim$9\%), with the remaining fraction consisting of
heavier nuclei such as C, N, O, Si, Mg, and Fe, as well as leptonic
and neutral components. At low energies (up to a few hundred MeV), the
proton flux within the heliosphere is significantly influenced by
solar activity\cite{pdg:2026aaa}. 

When primary cosmic rays enter the Earth's atmosphere, they interact
with atmospheric nuclei and initiate extensive air showers
(EAS). These hadronic interactions produce predominantly charged pions
($\pi^{\pm}$) and neutral pions ($\pi^{0}$). The $\pi^{0}$ mesons primarily
decay electromagnetically into two photons, contributing to the
electromagnetic component of the shower. Charged pions decay via weak
interactions, $\pi^{+}\rightarrow \mu^{+}\nu_{\mu}$ and
$\pi^{-}\rightarrow \mu^{-}\overline{\nu}_{\mu}$. Muons subsequently
decay with a mean lifetime of 2.2\,$\mu$s in the muon rest frame, producing electrons
(positrons) and neutrinos. Still, due to the large Lorentz factor/time
dilation, a large fraction of muons reach the Earth's surface before
they decay. Atmospheric muons therefore provide a
sensitive probe of primary cosmic-ray spectra and hadronic
interactions in the GeV--TeV energy range.    

The energy spectrum and composition of primary cosmic rays have been
extensively studied through direct measurements using balloon- and
satellite-borne experiments, as well as indirectly via surface-based
EAS arrays. Recent measurements of proton and helium spectra by
AMS-02\cite{ams2015_1,ams2015_2,ams2021_3},
CALET\cite{calet2022,calet2023} and DAMPE\cite{dampe2019, dampe2021,
dampe2026} have demonstrated significant deviations from a single
power-law behaviour. The maximum energy of the proton measured by the
AMS-02 and CALET experiments are 1.8\,TeV and 60\,TeV
respectively. DAMPE experiment is capable of measuring the primary
proton flux up to 100\,TeV. 
At energies above hundreds of TeV, direct
measurements are limited by detector acceptance and the steeply
falling spectrum. Indirect measurements using large-area surface
arrays extend the accessible range to larger energies region but
introduce systematic uncertainties arising from partial shower
sampling and intrinsic fluctuations in shower
development. Furthermore, their interpretation depends strongly on
hadronic interaction models, leading to additional model-dependent
uncertainties. 

Although the primary cosmic-ray flux is approximately isotropic, the
geomagnetic field imposes a rigidity-dependent 
cutoff that modifies the incident spectrum, also introducing the well-known
east–west asymmetry in secondary cosmic rays.
This effect is particularly
pronounced near the geomagnetic equator, where measurements are even rarer.
The zenith-angle distribution
of the atmospheric muons is commonly parameterised as
$I_{0}\cos^{n}\theta$, where $I_{0}$ represents the vertical flux and
$n$ characterizes the zenith angle dependence. The value of $I_{0}$ depends
strongly on the geomagnetic cutoff rigidity and atmospheric depth at
the observation site. Precise measurements of muon fluxes at different
geomagnetic latitudes also provide important constraints on atmospheric
neutrino flux calculations and reduce systematic uncertainties in
neutrino oscillation analyses. 

As part of the INO-ICAL R\&D programme, an RPC-only stack comprising
12 layers (without a magnetic field) was deployed at the IICHEP
Transit Campus. Using this setup, measurements of the vertical
integrated flux, angular exponent, and east--west asymmetry in
different zenith-angle bins were performed and compared with
phenomenological models\cite{pethu2017,pethu2020}. Subsequently, a magnetised
mini-ICAL configuration enabled the measurement of the muon charge
ratio in the momentum range 0.9--3\,GeV/c at a vertical cutoff rigidity of
17\,GV\cite{raj2025}. 

The present work reports measurements of charge-dependent atmospheric
muon spectra obtained with the magnetised mini-ICAL detector. The
experimental setup is described in Section~\ref{sec:expsetup}. The
reconstruction procedure and detector performance evaluation are
presented in Section~\ref{sec:data_analysis}. The Monte Carlo
framework, including CORSIKA-based air-shower generation, GEANT4
detector simulation, and event digitisation, is detailed in
Section~\ref{sec:mcsim}. The forward-folding methodology is discussed in
Section~\ref{sec:folding}. Section~\ref{sec:systerr} discusses the
different sources of systematic uncertainties on the observed
flux. The extracted fluxes and comparison with 
phenomenological models are presented in Section~\ref{sec:results},
followed by conclusions in Section~\ref{sec:concl}. 

\section{Experimental Setup}
\label{sec:expsetup}

\begin{figure}[htbp]
  \centering
  \includegraphics[width=0.9\textwidth]{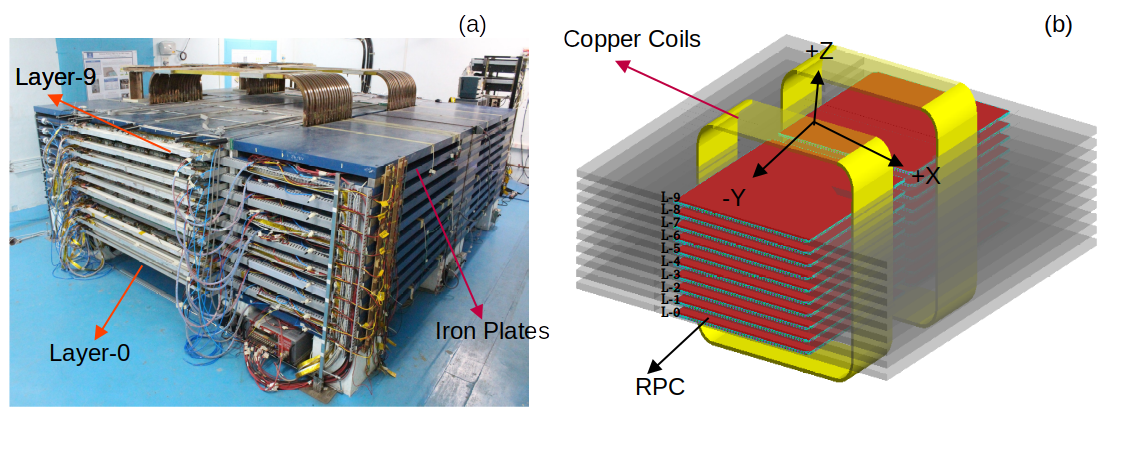}
  \caption{\label{fig:mini_ical} Photograph of the mini-ICAL detector stack and Geant4 modelling of
    the setup excluding the building.}
\end{figure}

The mini-ICAL detector is a scaled prototype of the ICAL design and
consists of 11 layers of low-carbon iron plates, each of dimension
4\,m\,$\times$\,4\,m\,$\times$\,0.056\,m, interleaved with RPCs as active
detector elements. The total mass of the stack is approximately
85\,tons. Resistive Plate Chambers (RPCs) of dimension 1.74\,m\,$\times$\,1.85\,m
are installed between successive iron layers, which act as muon detectors. 
The iron plates are magnetised by passing a current of 900\,A through
36 turns of copper coils, producing a magnetic field of approximately
1.5\,T in the central region (2\,m\,$\times$\,4\,m) of the detector along the -ve\,$y$-axis. The
magnetic field enables charge identification and momentum measurement
of traversing muons. 
Each RPC consists of two 3\,mm thick glass electrodes separated by a
2\,mm gas gap maintained using polycarbonate button spacers. The
chamber is hermetically sealed with edge spacers, except for two gas
inlets and two outlets for continuous gas circulation. The detector was
operated in avalanche mode using a gas mixture of Tetrafluoroethane
($\mathrm{C_{2}H_{2}F_{4}}$, 95.2\%), Isobutane
\mbox{(i-$\mathrm{C_{4}H_{10}}$}, 4.5\%), and Sulphur Hexafluoride
($\mathrm{SF_{6}}$, 0.3\%), supplied at a flow rate of
5\,sccm\footnote{SCCM = Standard Cubic Centimeters per Minute} per RPC chamber through
a closed-loop gas system\cite{kalmani2016}. 
A thin graphite coating is applied on the outer surfaces of the glass
electrodes to establish a uniform electric field across the gas
gap. High voltages of $\sim\pm$5\,kV are applied to the two
electrodes. Above the graphite layer, 100\,$\mu$m thick Mylar
insulation sheets are placed to isolate the pickup panel from the HV surface.
When a charged particle
traverses the gas volume, it initiates an avalanche, inducing signals
on copper pickup strips mounted on the panels. The pickup strips on
the two sides of the RPC are oriented orthogonally to provide
two-dimensional position information (X- and Y- coordinates). The strip
width and inter-strip gap are 28\,mm and 2\,mm, respectively. The total
number of strips on the X- and Y-planes is 58 and 61, respectively.
A side view and the GEANT4 modelling of the mini-ICAL stack are shown in
Fig.~\ref{fig:mini_ical} (a) and (b) respectively. 

The analog signals from the pickup strips are processed by NINO-based
preamplifiers and analog front-end boards\cite{ninoalice}, which
convert them into discriminated digital signals. These signals are
subsequently handled by FPGA-based RPCDAQ digital front-end boards\cite{mandar2022}.
The discriminated signals (with pulse widths extended to 100\,ns) from
all X- and Y-strips of an RPC are used to form independent pre-trigger
signals within the RPCDAQ.
These signals are transmitted to the Trigger Logic
Boards in the X (TLB-X) and Y (TLB-Y) planes independently.
Both TLB-X and TLB-Y generate coincidence triggers using 1-fold\footnote{1-fold triggers are formed by ORing any signal in the given layer.}
signals from four predefined layers (layers 6, 7, 8, and 9 for the data used in this analysis). The final
trigger module produces the global trigger by OR-ing the coincidences
from TLB-X and TLB-Y and distributes it to the RPCDAQ boards for data
latching. 
Upon receipt of a trigger, the strip hit data is latched in the
RPCDAQ system. Since the trigger path introduces a latency of
approximately 250\,ns, the strip hit\footnote{A hit is defined as an induced signal
in a strip exceeding a threshold of
100\,fC.} pulses are extended to take care of the delay in the trigger chain and its uncertainty.
During this data-taking period, the pulse width of the hit was extended 
up to 21\,$\mu$s. 
In addition to strip hit information, timing information was recorded
using multihit Time-to-Digital Converters (HPTDC 1190A), which register
the ORed signals from every eighth strip.

\section{Data Analysis}
\label{sec:data_analysis}
The muon data collected during the time period of December 12, 2018 to
December 28, 2018 was used for the analysis. The magnet was energised
during the day and de-energised at night.
The data analysis of the mini-ICAL experiment is based on two
categories of data:   
(i) non-magnetic data acquired during the night are used for detector calibration and
performance evaluation, while  
(ii) magnetic-field data recorded during the day are used for
charge-dependent muon measurements. The present study is based on
 $\sim$51 million triggered events recorded with the magnet
energised during stable operating conditions of the mini-ICAL
detector.

\subsection{Non-Magnetic Data}
\label{subsec:nonmagnetdata}

Data collected under non-magnetic conditions are used to determine
detector performance parameters, including pixel-wise (3\,cm\,$\times$\,3\,cm) efficiencies,
position and time offsets \& resolutions and hit multiplicity distributions.
To ensure stability of detector characteristics,
magnetic and non-magnetic data were recorded in alternating day–night
cycles so that environmental conditions remained comparable.
The reconstruction proceeds as follows. First, clusters are formed
independently in the X- and Y-planes by grouping consecutive strip
hits, with a maximum cluster size of three strips. Restricting the
cluster size suppresses contributions from shower activity, streamer pulses and
electronic noise. The cluster position is determined as the average
position of the constituent strips. For the calibration of each layer,
the track was reconstructed using all other layers, excluding the
layer under study.

The cluster positions in the XZ- and YZ-projections are fitted with straight lines,
\[
x = m_{xz} z + c_{xz}, \qquad
y = m_{yz} z + c_{yz},
\]
where $m_{xz}$, $m_{yz}$ and $c_{xz}$, $c_{yz}$ represent slopes and
intercepts in the respective projections. Events satisfying
$\chi^{2}/\mathrm{ndf} < 2$ and having hits in at least six layers are
selected for detector calibration studies. 

\begin{figure}[htbp]
\centering
\includegraphics[width=0.85\textwidth]{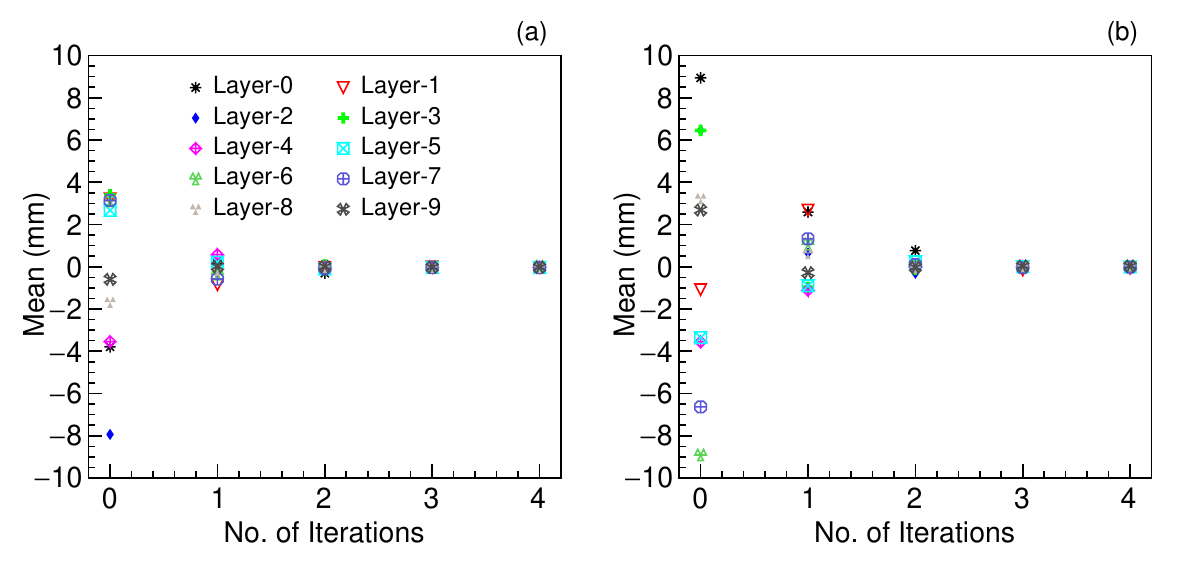}
\caption{\label{fig:poscorr} Mean (a) X-plane and (b) Y-plane
    position residuals of RPC layers after a few iterations.}
\end{figure}

Alignment corrections are determined using the selected events. These
include translational shifts in X- and Y- as well as relative tilts
between pickup panels. Position residuals in the X- and Y-plane are
computed, and the offsets are updated iteratively.
 The mean residuals for the X- and Y-planes are shown in
Figs.~\ref{fig:poscorr}(a) and (b), respectively.
After five iterations, a position
alignment accuracy better than 50\,$\mu$m is achieved for all
layers.

\begin{figure}[htbp]
\centering
\includegraphics[width=0.99\textwidth]{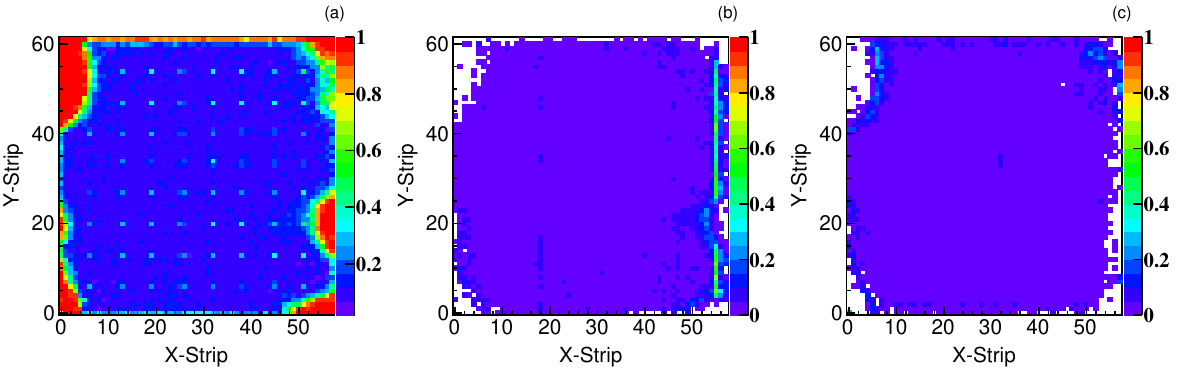}
\caption{\label{fig:correff} (a) Correlated inefficiency and uncorrelated inefficiencies in the (b) X-plane and (c) Y-plane.}
\end{figure}

The correlated inefficiency of a pixel is defined as the fraction of
selected events for which no hits are observed within one strip pitch
of the extrapolated position in both X- and Y-planes, relative to the
total number of events extrapolated to that pixel. This quantity is
sensitive to gas-gap related effects such as non-uniform gain or
spacer shadowing. 
Uncorrelated inefficiencies are evaluated independently for the X- and
Y- planes. For example, the X-plane uncorrelated inefficiency
corresponds to events with a valid hit in the Y-plane but not in the X-plane within one
strip pitch of the extrapolated position. These measurements identify
plane-specific effects such as inefficient or dead strips and
failure in front-end channels. The correlated and uncorrelated
inefficiencies in the X- and Y-planes are shown in
Figs.~\ref{fig:correff}(a), (b) and (c) respectively. The large
correlated inefficiency regions in the corners of the chambers are
mainly due to the variation in the gas gap due to side spacers placed
in the corners, thickness of glasses and nonuniformity in the resistive coating. 

In addition, hit multiplicity probabilities as a function of impact
position within a strip and intrinsic noise (excluding muon-induced
hits) are estimated. The detailed procedure for multiplicity
and noise estimation is described in J. M. John et al.~\cite{jim2023}.

In addition to the estimation of efficiencies and noise, the lead time
information recorded in the data for each RPC is utilised for the
localisation of the observed position based on the charge sharing
between the strips. The position residues as a function of the time
difference between the start strip and end strip for different
hit multiplicity was discussed in J. M. John et al.~\cite{jim2022}. With the use
of lead time correction, a substantial improvement was observed in the
position resolution of RPC at mini-ICAL, especially for hit multiplicity
of TWO. 


\subsection{Magnet Data}
\label{subsec:magnetdata}
\begin{figure}[htbp]
\centering
\includegraphics[width=0.99\textwidth]{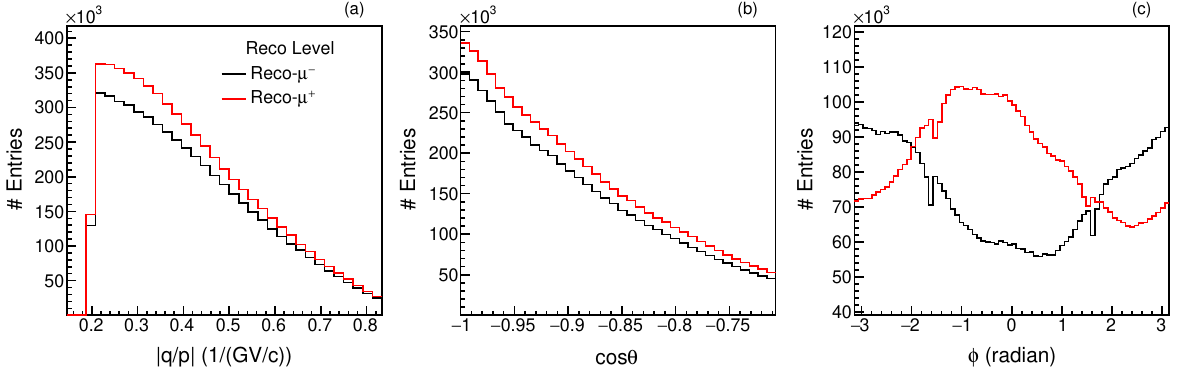}
\caption{\label{fig:datareco} The reconstructed distribution of
(a) charge/momentum ($q/p$),
(b) cosine of zenith angle ($cos\theta$) and (c) azimuthal angle
($\phi$) of muon.}
\end{figure}

Muon tracks with the magnetic field enabled are reconstructed
using the Kalman-Filter based algorithm\cite{kohlakal}. Before the
Kalman-Filter, the observed position and time information of a
single muon in the X- and Y-
planes are selected using the Track Finder algorithm. Since the data
are recorded for the large time window of 21\,$\mu$s with respect
to the muon trigger, a large number of noise hits could be associated along with
the muon tracks. The time information of the hits is utilised for the
preliminary filtering of valid muon hits. For this purpose, the
timing information of each RPC is obtained; the strip hit-time
within $\pm$\,50\,ns of the median of the time distribution is
considered for further analysis. Also, the average time delay of each layer with 
respect to the topmost RPC layer (i.e, layer 9, or layer 8 if layer 9 has
no hit or deviates by more than $\pm$50\,ns and so on) is computed. For
each event, the hits within $\pm$\,10\,ns from the average time delay
with respect to the topmost layer are selected for further analysis.
Selected hits from each layer in the X- and Y- planes are combined to form
two-dimensional (2D) hits, requiring the corrected time
difference between X- and Y-plane hits to be less than 7\,ns. 
Clusters are formed by grouping nearby 2D hits with a maximum hit
multiplicity of three in the X- and Y-planes, satisfying the timing
criterion of $|\Delta t|$\,<\,12\,ns.
Clusters from at least three out of five consecutive layers
are combined to form triplets, which are subsequently merged to
obtain complete tracks. 
Clusters belonging to identified tracks are processed using a Kalman
filter algorithm to estimate charge
and momentum. The state vector $(x, y, dx/dz, dy/dz, q/p)$ is
initialised using cluster positions and slopes derived from the topmost two
layers with valid signal, with an initial charge/momentum ($q/p$) set
to zero\,/(GV/c). Track propagation is
performed using the equation of motion of a charged particle in a
magnetic field. 
Measurement uncertainties arising from finite strip width, multiple
scattering, and propagation errors are incorporated into the
covariance matrix. State vectors are updated only if the extrapolated
position lies within $3\sigma$ of the measured cluster
position. Forward and backward filtering iterations are performed to
smooth the track parameters. 
The extracted parameters charge/momentum ($q/p$), zenith angle
($\theta$), and azimuthal angle ($\phi$) are
determined at the topmost RPC layer. These parameters are then
extrapolated to the roof of the detector hall, accounting for energy loss and
multiple scattering in the overburden materials. The reconstructed
events with a minimum of 8 layers, fit probability greater than 10$^{-5}$,
the reconstructed 0.2<$|q/p|$<0.833\,/(GV/c) (1.2\,$<\,|p|\,<$\,5\,GeV/c) and
$cos\theta$<$-$0.707 ($\theta$>135$^{\circ}$) are considered for the
further process. The distribution of reconstructed 
$q/p$, $cos\theta$ and $\phi$ for the data are shown in
Figs.~\ref{fig:datareco}(a), (b) and (c) respectively.
Here the muon tracks along the -ve\,$z$-axis is represented as zenith
angle, $\theta$\,=\,180$^{\circ}$ ($cos\theta$\,=\,$-$1) and the azimuth
angle $\phi$\,=\,0 radian represent the +ve\,$x$-axis of the mini-ICAL detector.
There are kinks near $\phi=\pm\pi/2$ radian, which arise primarily
due to the finite width of the strip. In the present case, the
magnetic field in the central region of mini-ICAL is directed along the -ve\,$y$-axis, corresponding to
$\phi=-\pi/2$ radian. 

\section{Monte-Carlo Simulation}
\label{sec:mcsim}
The Monte-Carlo framework consists of three components: generation of
cosmic-ray induced air showers using CORSIKA, detector simulation
using GEANT4, and subsequent digitisation of detector response. 

\subsection{CORSIKA simulation}
\label{sec:corsika}
Extensive air showers are simulated using the CORSIKA package
(v7.6300) \cite{corsika}. The simulation incorporates primary particle
type, input energy spectrum, geomagnetic field configuration, site
altitude, and direction-dependent rigidity cutoff corresponding to the
experimental location. US Standard atmosphere with the ``CURVED''
atmosphere option is used in the simulation.
The geomagnetic field components at the observation site are $B_{x} =
40.431\,\mu$T (horizontal component) and $B_{z} = 4.705\,\mu$T
(vertical downward component), obtained from the GEOMAG model
\cite{geomag}. The directional rigidity cutoff as a function of
$(\theta,\phi)$ is calculated using the back-tracing technique based
on the IGRF-12 geomagnetic field model. The effect of the
solar magnetic field on the incident proton and helium spectra was not
explicitly included in the CORSIKA simulations. However, the primary
cosmic-ray spectrum used in the simulation starts at 10 GV, while the
geomagnetic cutoff rigidity at the mini-ICAL site is approximately
17\,GV. Therefore, the primary particles contributing significantly to the
measured muon flux are predominantly above the energy range where
solar modulation is expected to have a strong effect.
In addition, temporal variations of the muon flux associated with
solar activity were not included in the simulation. Such variations
are expected to be small compared with the experimental uncertainties
in the present measurement; measurements with the GRAPES-3 muon
telescope have reported variations of less than 1\% in the relevant
context \cite{kojima_grapes3}. 
A total of $2\times10^{7}$ primary protons and $2\times10^{6}$ helium
nuclei are generated with energies between 10\,GeV and 1\,PeV. The
contribution of other elements is found to be very small; thus, we neglect their contribution. The
primaries are sampled uniformly in solid angle over the ranges
$0^\circ \leq \theta \leq 85^\circ$ and $0^\circ \leq \phi <
360^\circ$, with a spectral index fixed at $-2.7$. 
Secondary particles are recorded at the observation level
corresponding to the experimental site. The minimum kinetic energy
thresholds for storing secondaries are 0.1\,GeV for muons and hadrons,
and 0.5\,GeV for electrons/positrons and photons. Since the secondary
particle distributions at ground level depend on hadronic interaction
modelling, FLUKA, GHEISHA and URQMD are used for low-energy interactions($E < 80$\,GeV)
and SIBYLL for high-energy interactions ($E \geq 80$\,GeV). Given that
the present measurement focuses on GeV-scale muons, the results are
primarily sensitive to the low-energy hadronic model\cite{raj2025}.
The zenith angle $\theta$ of the particle track is defined as the angle of the particle
momentum vector with respect to the
-ve\,$z$-axis; $\phi = 0^\circ$ corresponds to geographic North, and
$\phi = 90^\circ$ corresponds to East. The generated secondaries are
stored in three-dimensional distributions of $q/p$, $\cos\theta$, and
$\phi$ after translating into the detector coordinate system, which serves as inputs to the detector simulation. 

\subsection{GEANT4 simulation and digitisation}

\begin{figure}[htbp]
  \centering
  \includegraphics[width=0.99\textwidth]{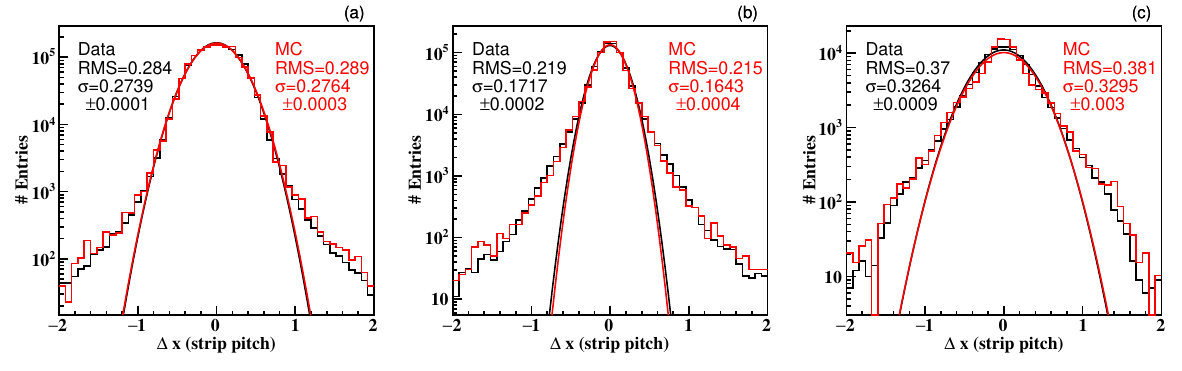}
  \caption{\label{fig:posreso} Comparison of position residual
    distributions between data and Monte-Carlo simulation for hit
    multiplicities (a) ONE, (b) TWO, and (c) THREE.}
\end{figure}

\begin{figure}[htbp]
  \centering
  \includegraphics[width=0.99\textwidth]{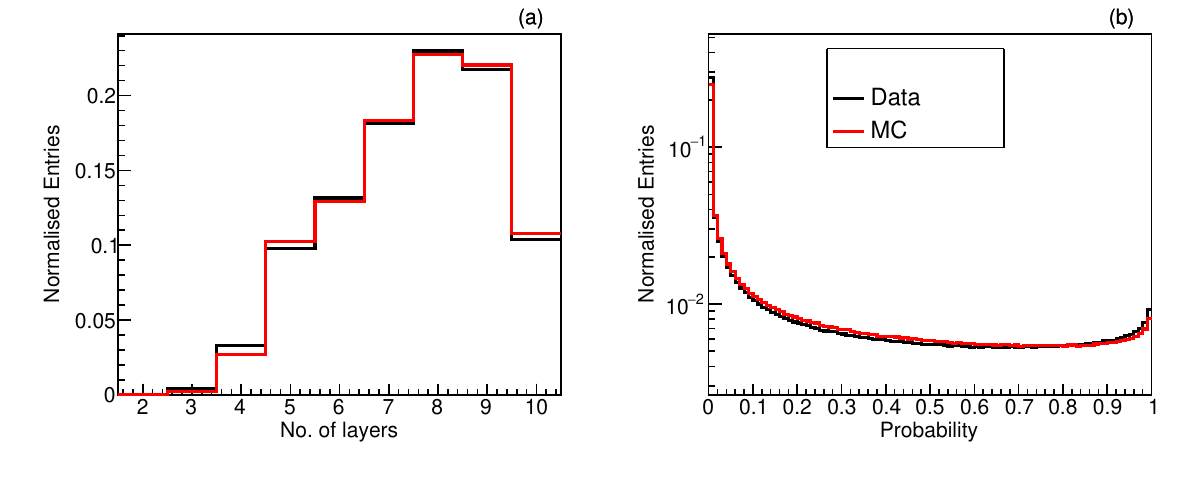}
  \caption{\label{fig:probndf}  Comparison of data and Monte-Carlo
    simulation for (a) number-of-layer used in the fit
    and (b) fit probability.} 
\end{figure}

\begin{figure}[htbp]
  \centering
  \includegraphics[width=0.6\linewidth]{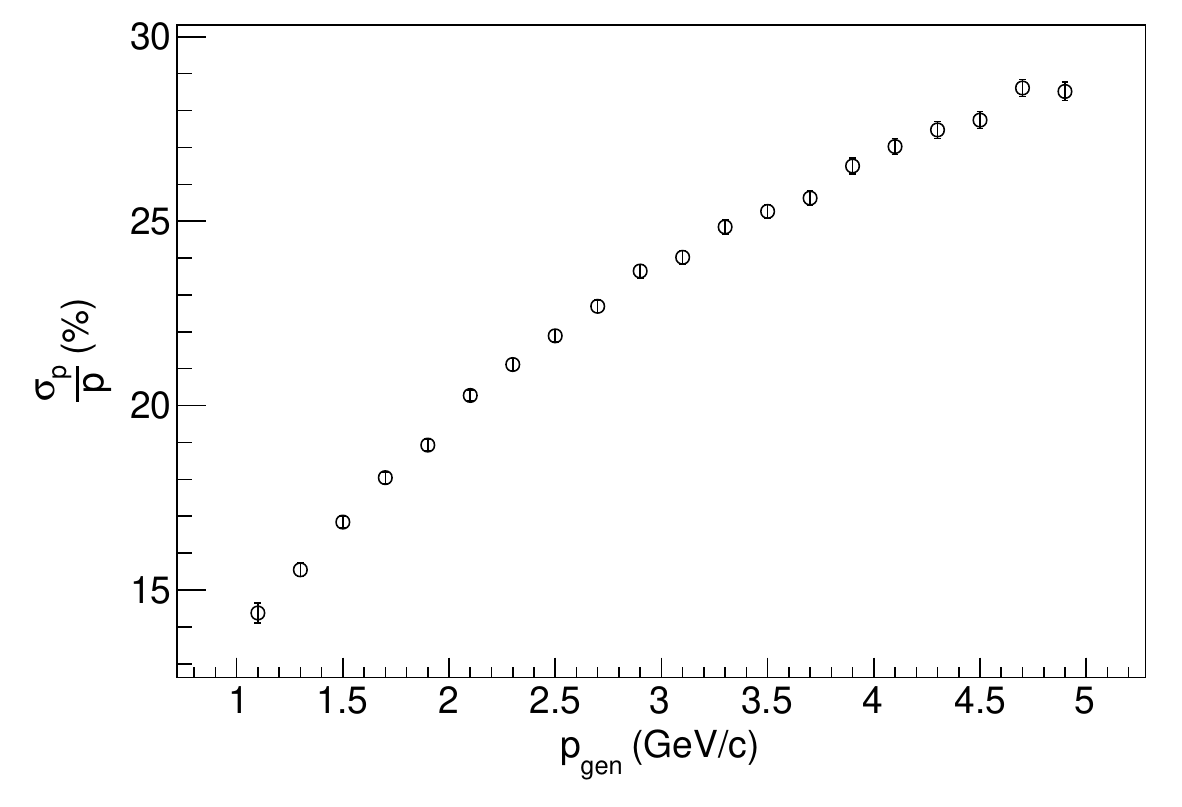}
  \caption{\label{fig:momreso} The momentum resolution (\%) as a function of muon momentum.}
\end{figure}

The secondary particles generated by CORSIKA are propagated through a
detailed GEANT4-based detector model\cite{geant4}, which includes the
mini-ICAL geometry and the experimental hall. 
The simulation and digitisation procedure consists of the following steps:
(i) For each secondary particle, the position $(x,y)$ is sampled
uniformly over the top trigger layer (layer 9). The particle identity,
$q/p$, $\cos\theta$, and $\phi$ are assigned from the
CORSIKA-generated distributions. 
(ii) The trajectory is extrapolated to the bottom trigger layer (layer
6). Only particles satisfying the geometrical acceptance criteria with
2\,\% increase in detector size to accommodate the effect of multiple scattering and bending due to magnetic field are
retained. These tracks are then extrapolated back to the detector roof, and
propagation through the detector volume is simulated including
magnetic field effects. 
(iii) The simulated energy deposits in RPC layers are converted into
strip hits by digitising the signals according to the strip width in
both X- and Y- planes. 
(iv) To reproduce realistic detector response, layer-dependent
efficiencies, noise rates, and timing offsets derived from
non-magnetic calibration data (Section~\ref{sec:data_analysis}) are
incorporated during digitisation.
(v) Since the simulation does not incorporate the avalanche process,
the lead time correction as a function of position residues from
Non-Magnet data is incorporated during the digitisation.
The lead time generally represents the time at which the signal crosses
the fixed set threshold, which strongly depends on the gain of the
avalanche produced while muon passes through the RPC gas-gap as well as
shared induced signal in nearby strips. It is
possible to locate the position of the muon in a strip due to sharing of
the signal between nearby strips, if the hit multiplicity is more than
one.
The position residues of the non-Magnetic MC are simulated and digitised to verify
the lead time correction. The comparison of position residues of
non-Magnet data and non-Magnet MC is shown after incorporating the
lead time correction for hit multiplicity one (no lead time correction), two and three are shown in
Figs.~\ref{fig:posreso}(a), (b) and (c) respectively.    
The digitised Monte-Carlo events are processed using the same
reconstruction algorithm as applied to experimental data.
Comparisons
between data and simulation for key observables, the number of layers
and fit probability are
shown in Figs.~\ref{fig:probndf}(a) and (b), respectively.
The momentum resolution of the mini-ICAL detector was obtained from the
simulated Monte Carlo (MC) using the detector parameters
described above. Due to the saturation of reconstructed momentum,
the $\Delta p$ distributions have an
asymmetric shape. Thus, the distributions are fitted with the
asymmetric Gaussian, and the average of two $\sigma$s is shown in
Fig.~\ref{fig:momreso}.
The momentum
resolution deteriorates with increasing momentum, primarily
due to the finite position resolution of the detector.
 Thus, this analysis restricted the reconstructed momentum up to 5\,GeV.
In MC, the
GEANT4 simulation and digitisation were performed considering the
width of the pickup strips is uniform along the entire strip length. But in reality, the non-uniformity in
the pickup strip pitch, bending of the strips and uncertainty in the
position-dependent multiplicity would lead to poor momentum
reconstruction in data compared to the MC. To verify that, the
reconstruction was performed using the hits present in odd and even
layers using the data and MC. The distribution of difference between $(q/p)_{odd}$ and
$(q/p)_{even}$ in MC was found to be better than the data. So the
reconstructed $q/p$ in MC is smeared by
$\sigma_{smear}^{q/p}$=0.08\,/(GV/c) to match data and MC.     

\section{Forward-Folding Method}
\label{sec:folding}

The observed particle spectra are distorted by detector effects,
including multiple scattering, energy loss, finite resolution,
detection efficiency, and statistical fluctuations. The underlying
true distribution can be inferred using either (i) unfolding
techniques, which invert the detector response matrix to recover the
true spectrum, or (ii) forward-folding methods, where a model of the
true distribution is propagated through the detector response and
compared directly with the measured data. 
Unfolding methods are often sensitive to statistical fluctuations and
can amplify small uncertainties in the measured spectrum, particularly
in regions of limited resolution\cite{unfold1, unfold2, folding}.
Given the moderate momentum resolution of $\sim$\,30\,\% at 5\,GeV/c in
the mini-ICAL detector, an iterative forward-folding approach is
adopted in this work. The true momentum spectrum from CORSIKA is considered in the
range $|p| \geq 0.71$\,GeV/$c$, with $130^\circ \leq \theta \leq 180^\circ$
and full azimuthal coverage. 
Events at large zenith angles suffer significant multiple scattering
in the iron plates, leading to degraded momentum and azimuthal
reconstruction. Therefore, reconstructed events with $\cos\theta$ >
$-$0.707 ($\theta$ < 135$^\circ$) are excluded from the folding analysis. 

\subsection*{Response matrix}

\begin{figure}[htbp]
\centering
\includegraphics[width=0.85\linewidth]{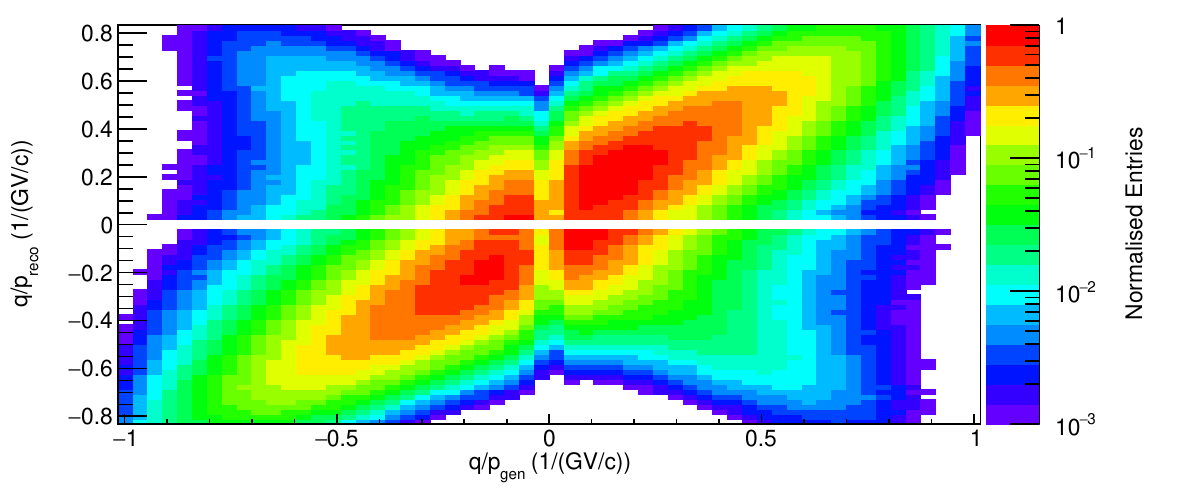}
\caption{\label{fig:qbypmatrix} Response matrix for the $q/p$ distribution.}
\end{figure}

\begin{figure}[htbp]
\centering
\includegraphics[width=0.99\textwidth]{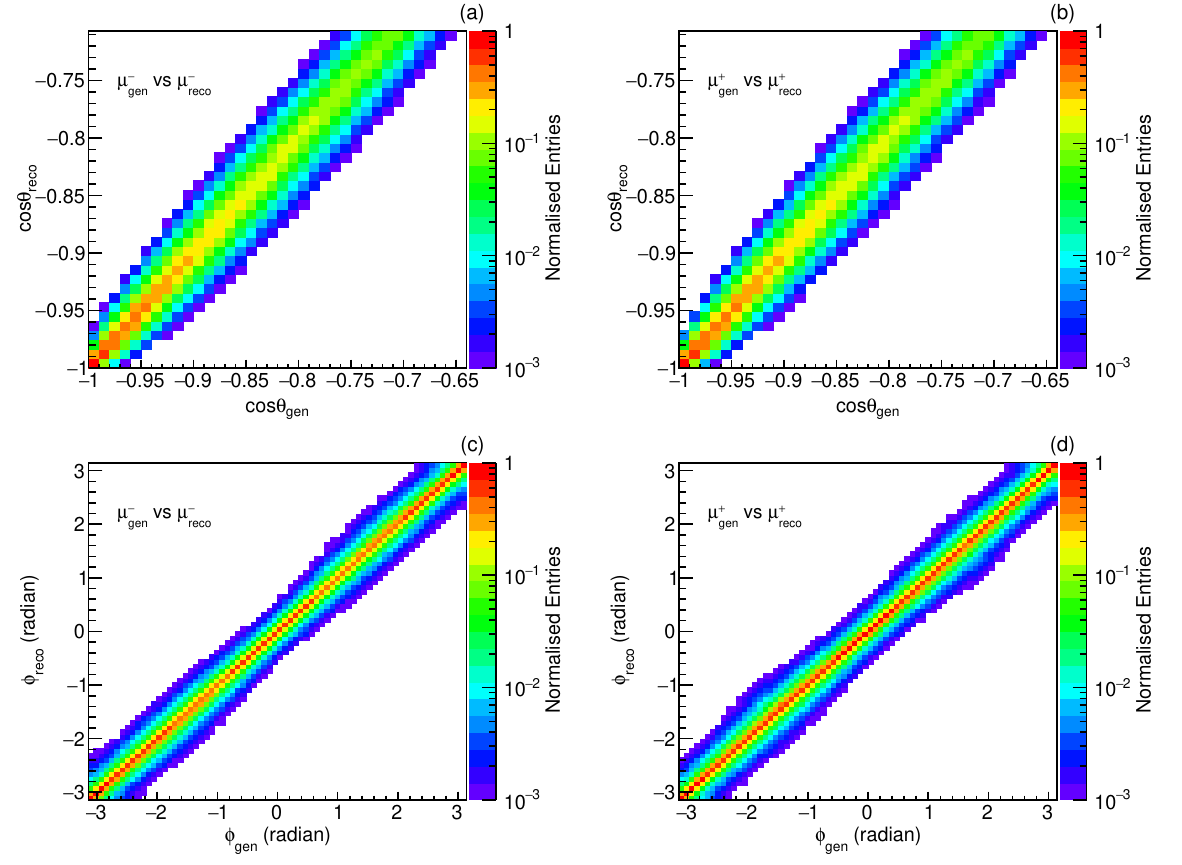}
\caption{\label{fig:costhephimatrix} (a) and (b) are response matrices for
$\cos\theta$ and (c) and (d) are $\phi$ for generated and
  reconstructed $\mu^{-}$ and $\mu^{+}$ events.} 
\end{figure}

\begin{figure}[htbp]
\centering
\includegraphics[width=0.99\textwidth]{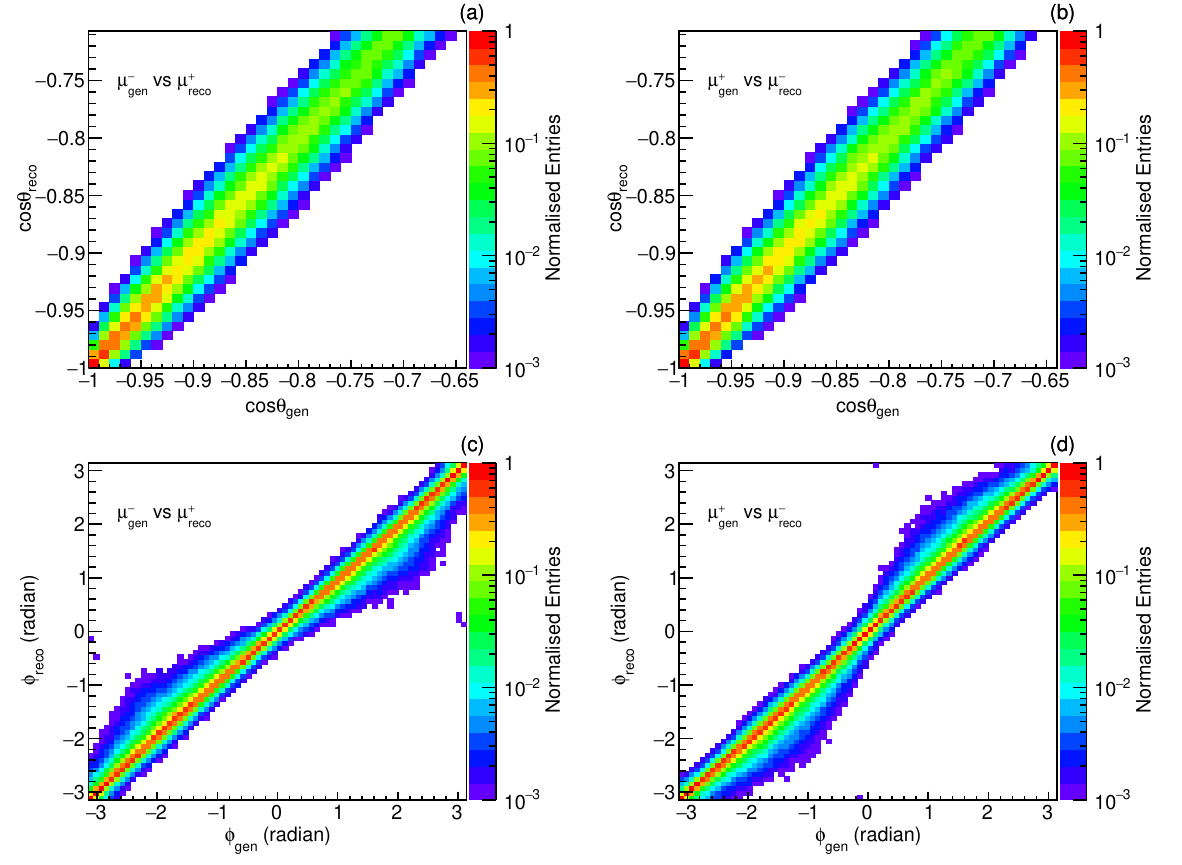}
\caption{\label{fig:costhephimatrix_2} (a) and (b) are response matrices for
  $\cos\theta$ and (c) and (d) for $\phi$ for events with reconstructed
momentum of opposite charge.} 
\end{figure}

The detector response matrix is derived from Monte-Carlo simulations
and is constructed as a multidimensional mapping between generated and
reconstructed variables $(q/p, \cos\theta, \phi)$. Since these
kinematic variables are correlated, the response matrix is evaluated
in three dimensions. The response matrix could be obtained from the
MC samples simulated with input cosmic muon spectrum for three
different models, FLUKA, GHEISHA and URQMD. For this study, all the MC
samples from all three models are combined to obtain a single
response matrix used for the forward folding.
The response matrix for $q/p_{\mathrm{gen}}$ versus
$q/p_{\mathrm{reco}}$ is shown in Fig.~\ref{fig:qbypmatrix}. For large
$|q/p_{\mathrm{gen}}|$ (low momentum), the response is approximately
diagonal. 
For smaller
$|q/p_{\mathrm{gen}}|$ (higher momentum), the reconstructed values
become broadly distributed due to limited detector resolution. 
The response matrices for $\cos\theta$ and $\phi$ for correctly
reconstructed $\mu^{-}$ and $\mu^{+}$ events are shown in
Fig.~\ref{fig:costhephimatrix} (a) to (d). Charge misidentified response matrix
for $\cos\theta$ and $\phi$, corresponding to events generated as $\mu^{-}$
($\mu^{+}$) and reconstructed as $\mu^{+}$ ($\mu^{-}$), are presented
in Fig.~\ref{fig:costhephimatrix_2} (a) to (d). 

\subsection*{Iterative forward-folding procedure}

The forward-folding procedure consists of three main steps.

\paragraph{1) Parameterization of the true spectrum:}

\begin{figure}[htbp]
\centering
\includegraphics[width=0.99\textwidth]{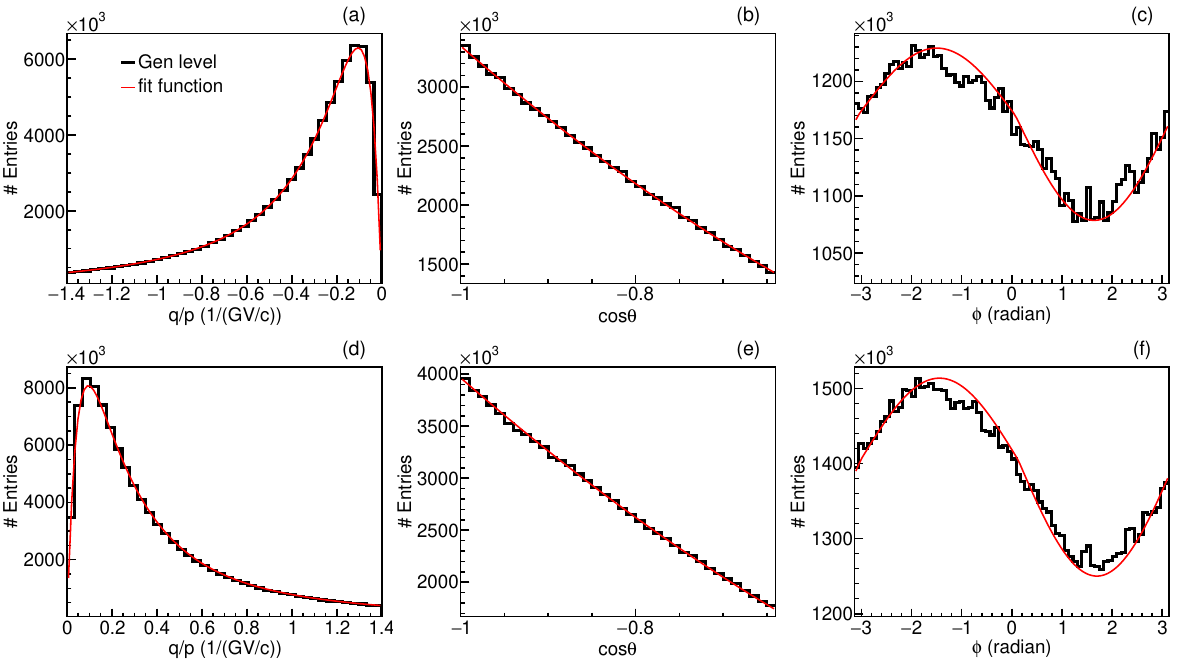}
\caption{\label{fig:gendist}  (a), (b) and (c) are generated-level distributions of $q/p$, $\cos\theta$, and $\phi$ for $\mu^{-}$ and (d), (e) and (f) are for $\mu^{+}$. Details of the fit functions are given in the text.}
\end{figure}

The generated $(q/p, \cos\theta, \phi)$ distributions obtained from
CORSIKA are parameterised using analytic functions. The $q/p$
distribution is modelled using a functional form analogous to capacitor
charging-discharging behaviour, while $\cos\theta$ is described by a
power-law function and $\phi$ by a sum of two sinusoidal functions. Separate
parameterizations are obtained for $\mu^{-}$ and $\mu^{+}$ which are shown in  
Fig.~\ref{fig:gendist} (a) to (f).  

\paragraph{2) Weighting of the true distribution:}

The true spectrum is modified by introducing weights
\[
\omega(q/p), \quad \omega(\cos\theta), \quad \omega(\phi),
\]
obtained by varying the fit parameters around their nominal values.

The initial normalization for $\omega(q/p)$ is defined as
\[
\omega(q/p)_{\mu^{\pm}} =
\frac{T_{\mathrm{total}} \times N^{rate}_{\mathrm{Honda}}(\mu^{\pm})}
     {N_{\mathrm{gen}}^{\mathrm{MC}}(\mu^{\pm})},
\]
where $T_{\mathrm{total}}$ is the total data-taking period,
$N^{rate}_{\mathrm{Honda}}$\footnote{The Honda flux was obtained by private
communication.} is the expected muon rate (Hz) at the INO site based
on Honda flux predictions, and $N_{\mathrm{gen}}^{\mathrm{MC}}$ is the
number of generated Monte-Carlo muons. The initial weights for
$\cos\theta$ and $\phi$ are set to unity. 

The total weight applied to the generated spectrum is
\[
W = \omega(q/p)\,\omega(\cos\theta)\,\omega(\phi).
\]

The weighted true spectrum is then multiplied by the total efficiency,
\[
\epsilon_{\mathrm{total}} =
\epsilon_{\mathrm{geom}} \times
\epsilon_{\mathrm{trigger}} \times
\epsilon_{\mathrm{selec}},
\]
where each efficiency component is evaluated as a function of $(q/p, \cos\theta, \phi)$.

\paragraph{3) Folding and $\chi^2$ minimization:}

\begin{figure}[htbp]
\centering
\includegraphics[width=0.99\textwidth]{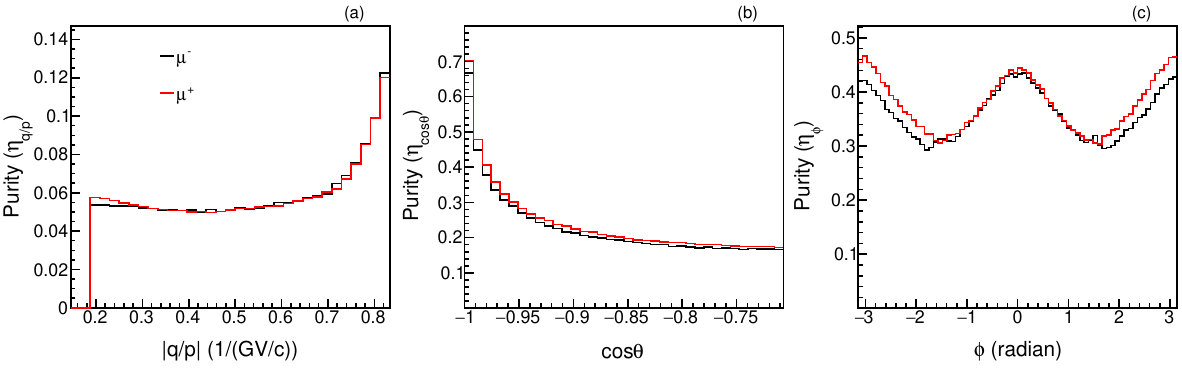}
\caption{\label{fig:purity} (a), (b) and (c) are reconstruction purity as a function of
$|q/p|$, $\cos\theta$, and $\phi$ respectively.} 
\end{figure}

The weighted true distribution is folded with the response matrix to
obtain the predicted reconstructed distribution. The agreement with
data is quantified using a $\chi^2$ statistic defined as 
\begin{equation}
\chi^{2}(x) =
\sum_{i=1}^{N_{\mathrm{bins}}}
\frac{\left[N^{\mathrm{data}}_{i}(x) - N^{\mathrm{folded}}_{i}(x)\right]^2}
     {N^{\mathrm{data}}_{i}(x)}
\,\eta_i(x),
\label{eqn:chisq}
\end{equation}
where $x \in \{|q/p|, \cos\theta, \phi\}$, and $\eta_i(x)$ represents
the purity in bin $i$, defined as the fraction of events generated and
reconstructed in the same bin relative to total reconstructed events in
that bin. The purity of $|q/p|$, $\cos \theta$ and $\phi$ are shown in Fig.~\ref{fig:purity} (a), (b) and (c) respectively. The purity of the events increases and reaches a maximum of 0.12 for
$|q/p|$>0.7 /(GV/c) ($\sim|p|$ < 1.4\,GeV/c) due to reasonable bending of muons in
the mini-ICAL. The low value of the purity in $q/p$ is primarily due to the
finite resolution and consequently the bin migration. The events with $|q/p|$<0.5
also little increases due to
the increase in the width of the bin in momentum space. 
The purity as a function of
$\cos\theta$ decreases with increasing $\cos\theta$ of the muon tracks
due to the decreasing bin width in the $\theta$-coordinate and
multiple scattering at large zenith angle in the mini-ICAL.
The reconstruction of $\phi$ is dominantly affected due to the combination of
strip-type detector geometry and the direction of the 
magnetic field. The muons at $\phi = \pm\,\pi/2$ radian (along and
opposite to the magnetic field direction) is poorly
reconstructed, which results in a decrease in the purity around the
$\phi = \pm\,\pi/2$ radian region.
Due to the poor position resolution and the small lever arm of
mini-ICAL, migration between $q/p$ bins is significant, particularly
at large momentum (small $|q/p|$). Consequently, the $|q/p|$ purity is
lower than the purity obtained for the angular variables. 
The angular resolutions are considerably better, with
typical resolutions of $\sigma_{\cos\theta}$ $\sim$0.0267 and
$\sigma_{\phi}$ $\sim$0.0949 radian,  which are comparable to the
corresponding bin widths, 0.0081 and 0.0873 radian in $\cos\theta$ and
$\phi$ distributions respectively,
resulting in smaller bin-to-bin
migration and hence relatively higher purities.

\begin{figure}[htbp]
\centering
\includegraphics[width=0.99\textwidth]{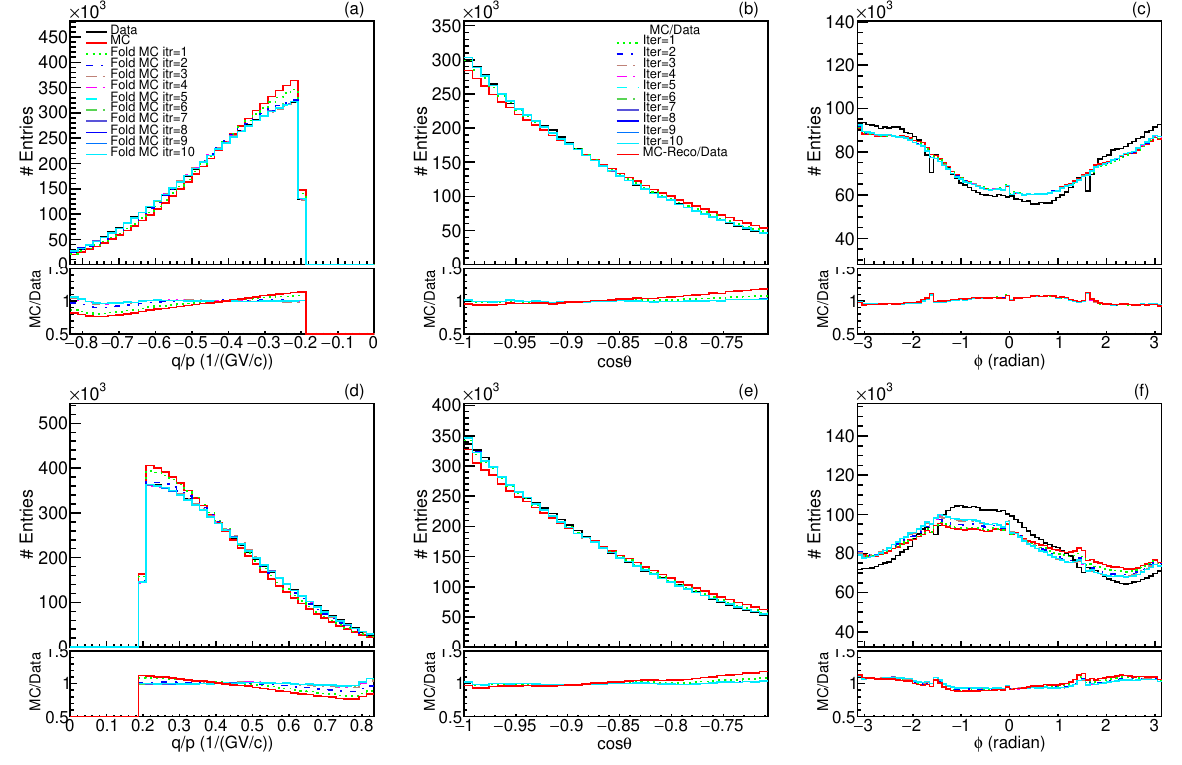}
\caption{\label{fig:folddata} (a), (b) and (c) are the comparison of reconstructed and folded MC with data for $\mu^{-}$ and (d), (e) and (f) are for $\mu^{+}$ of $q/p$, $\cos\theta$, and $\phi$ respectively.} 
\end{figure}

\begin{figure}[htbp]
\centering
\includegraphics[width=0.99\textwidth]{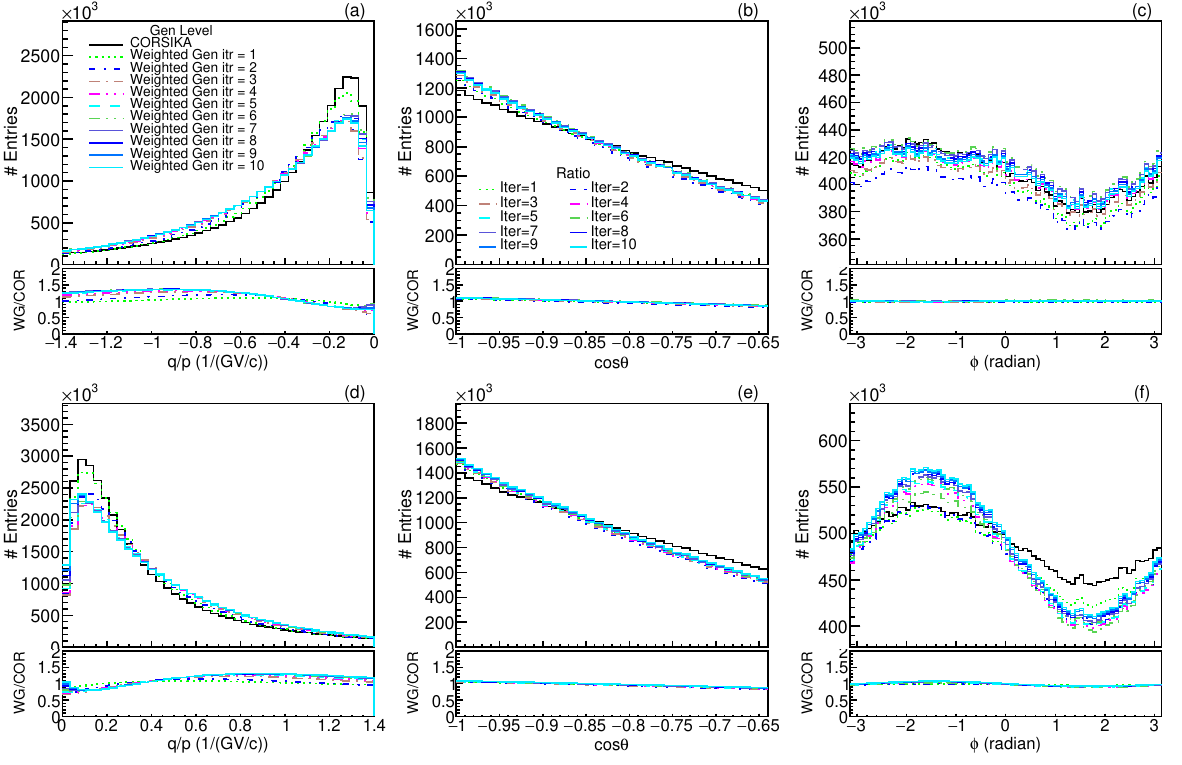}
\caption{\label{fig:weightgen} (a), (b) and (c) are the weighted generated-level distributions for $\mu^{-}$ and (d), (e) and (f) are for $\mu^{+}$  as functions of $q/p$, $\cos\theta$,
and $\phi$ from the iterative method respectively.} 
\end{figure}

The fit parameters are iteratively updated by minimising
$\chi^2$ for each kinematic variable in sequence: $q/p$, followed by $\cos\theta$ and $\phi$ for
$\mu^{-}$ and $\mu^{+}$ together. The procedure is repeated until convergence
is achieved. 
The comparison between folded spectra and reconstructed data for
$\mu^{-}$ and $\mu^{+}$ in
successive iterations as a function of $q/p$, $\cos\theta$ and $\phi$
is shown in Fig.~\ref{fig:folddata}(a) to (f). Convergence
is typically reached after three iterations.
Since the folded distributions in $q/p$, $\cos\theta$, and $\phi$
become more consistent with the reconstructed data after the fourth
iteration, the weighted CORSIKA-level true distributions obtained
beyond the fourth iteration are expected to provide a better
representation of the true cosmic-muon distributions at the
experimental site. The comparison of the weighted CORSIKA-level true
distributions of $q/p$, $\cos\theta$, and $\phi$ at different
iterations, together with the corresponding unweighted CORSIKA-level
distributions for $\mu^{-}$ and $\mu^{+}$, are shown in
Figs.~\ref{fig:weightgen}(a)--(c) and ~\ref{fig:weightgen}(d)--(f), respectively. 

Comparing the weighted CORSIKA-level $q/p$ distribution at the tenth
iteration with the corresponding unweighted distribution, a
significant difference in the spectral shape is observed. Similarly,
for $\cos\theta$, the weighted distributions exhibit a steeper fall
compared with the unweighted distributions for both $\mu^{-}$ and
$\mu^{+}$. In the case of the $\phi$ distribution, the east--west
asymmetry in the weighted distribution for $\mu^{+}$ is more
pronounced than that observed in the unweighted distribution. 


\section{Systematic Uncertainties}
\label{sec:systerr}
The obtained results are affected by several detector-related
systematic uncertainties. Therefore, various systematic studies were
performed to evaluate their impact on the central value of the
weighted generated-level spectrum. The systematic variations
considered in this study are as follows:

\begin{itemize}
\item Since the mini-ICAL detector is highly sensitive to low-energy
muons, the material budget of the experimental hall can affect the
measured results. To evaluate this effect, the roof thickness in the
detector hall in simulation was varied by its estimated uncertainty of
$\pm$\,10\,$\%$. 
\item The magnetic field map used in the simulation has an uncertainty of $\sim$3.3\,$\%$\cite{honey}. So the magnetic field in the mini-ICAL increased and decreased by 3.3\,$\%$.
\item Apart from the experimental hall, the iron plates constitute the
largest source of material in the mini-ICAL detector. According to the
vendor specifications, the plate thickness varies between 54.88\,mm and
57.12\,mm across the detector layers. To evaluate the corresponding
systematic uncertainty, the iron plate thickness was varied within
this specified range in the simulation. 
\item In addition to the detector modelling, the pixel-wise trigger efficiencies used during digitisation were varied by $\pm$\,1$\sigma$ to evaluate their impact on the measured central flux.
\item The reconstructed events are selected with a minimum number of layers $\geq$\,7 and $\geq$\,9 used to estimate the muon flux.
\item The $\sigma_{smear}^{q/p}$ of MC increased and decreased to its
uncertainty, 0.01\,/(GV/c) from the central value.
\item The position alignment in the data is obtained with less than 50\,$\mu$m precision in most of the layers; the obtained precisions of different layers from data were added during the digitisation to see any bias in the momentum spectrum.
\end{itemize}
For convenience 
in comparison with the other experimental results in the further sections, the q/p
distribution is converted to a momentum distribution. The zenith angle
distribution ($\cos\theta$) is translated in such a way that the vertically
downgoing muons (along -ve\,$z$-axis) are represented as $\theta^{\prime}$=0$^{\circ}$ (or
$\cos\theta^{\prime}$=1). Similarly, the arrival direction of the azimuthal angle
$\phi^{\prime}$=0 radian (towards the magnetic North) and
$\phi^{\prime}$=$\pi$/2 radian (towards the magnetic West). 

\begin{figure}[htbp]
\centering
\includegraphics[width=0.99\linewidth]{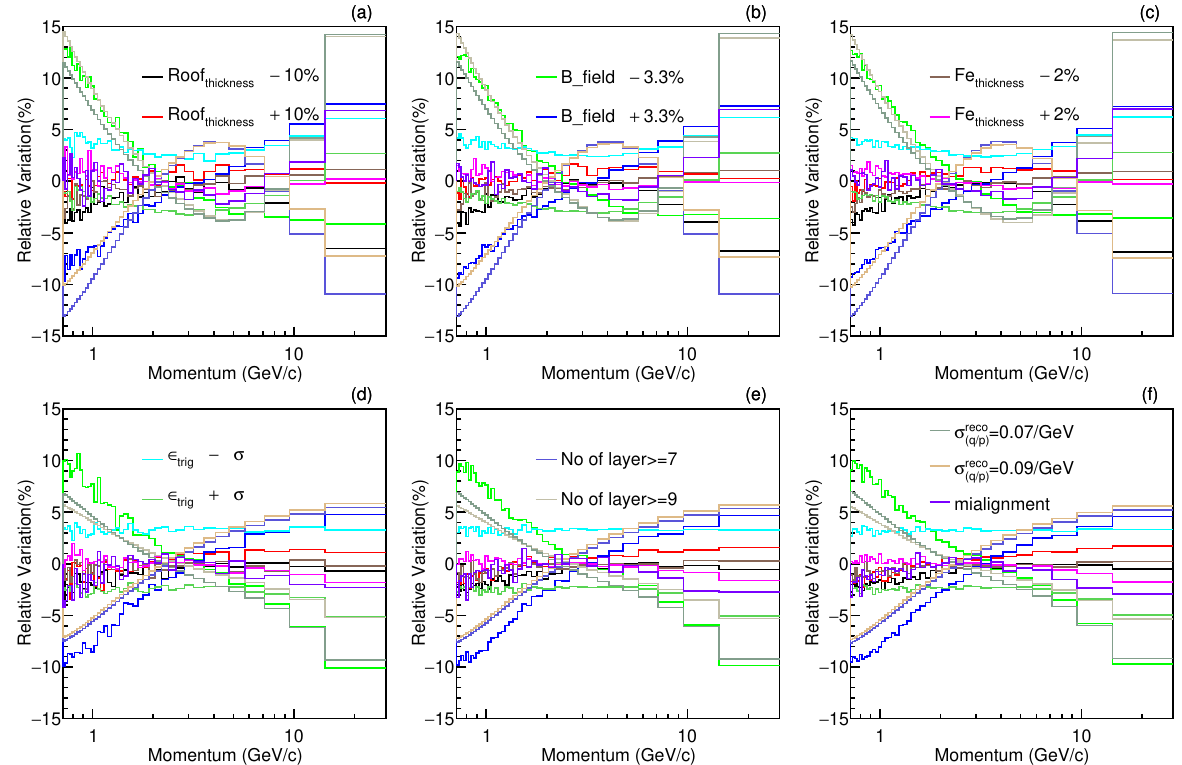}
\caption{\label{fig:fluxmomsyst} (a), (b) and (c) are systematic
uncertainties on the momentum spectra of cosmic muons for zenith angle
acceptance $\cos\theta^{\prime}$ > 0.98, 0.94 and >0.90 respectively
for $\mu^{-}$ and (d), (e) and (f) are the same for $\mu^{+}$.} 
\end{figure}

\begin{figure}[htbp]
\centering
\includegraphics[width=0.99\linewidth]{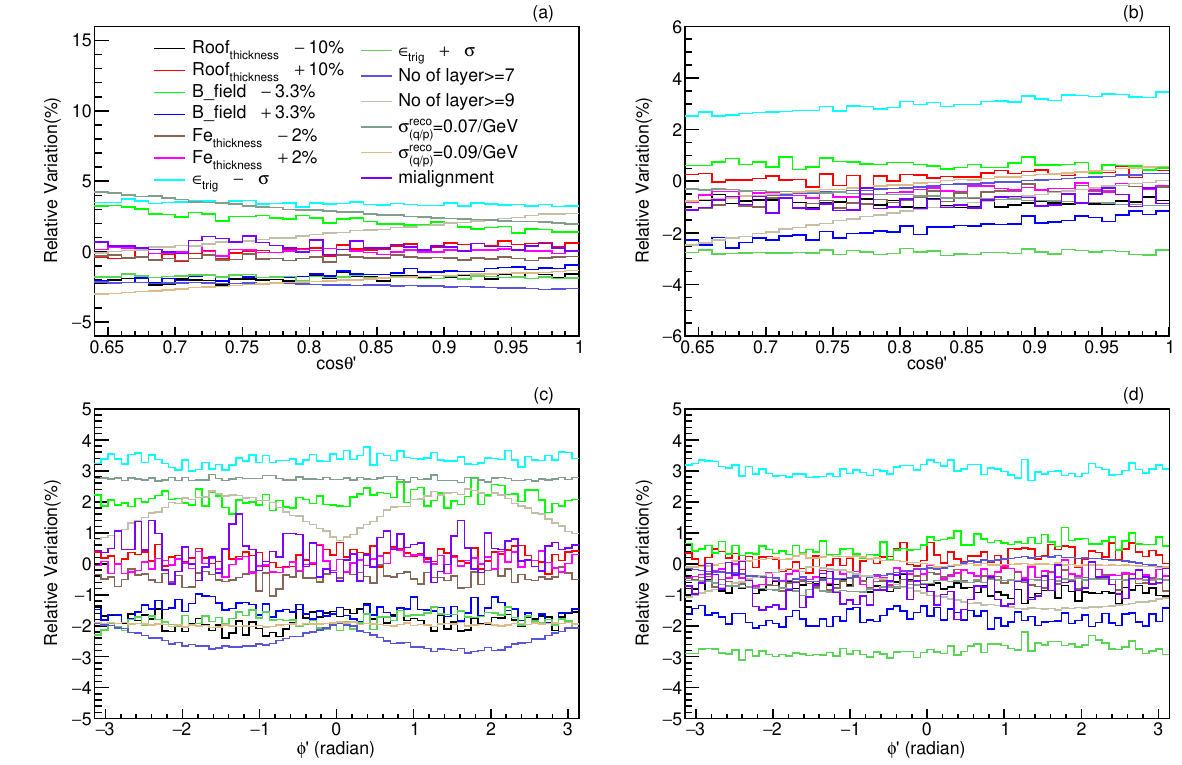}
\caption{\label{fig:fluxthetasyst} (a) and (b) are the systematic
errors on the Zenith-angle spectra of cosmic muons for (a) $\mu^{-}$
and (b) $\mu^{+}$ respectively. (c) and (d) are the systematic errors
on azimuthal spectra for $\mu^{-}$ and $\mu^{+}$ respectively. } 
\end{figure}

The relative variation of the muon flux from the central value as
function of momentum for different zenith angle acceptance
$\cos\theta^{\prime}$>0.98 ($\theta^{\prime}$<11.48$^{\circ}$),
$\cos\theta^{\prime}$>0.94 ($\theta^{\prime}$<19.95$^{\circ}$) and
$\cos\theta^{\prime}$>0.90 ($\theta$<25.84$^{\circ}$) are shown in
Fig~\ref{fig:fluxmomsyst} (a) to (f), where (a) to (c) are for
$\mu^{-}$ and (d) to (f) for $\mu^{+}$ respectively. 
The uncertainties on the measured momentum flux due to the material budget are less than 5\,$\%$. The major uncertainties are contributed by the variation in the magnetic field map in the simulation, change in the trigger efficiency during the digitisation, selection criteria based on the number of layers and the smearing in the q/p distribution in MC. Hence, the overall uncertainty on the momentum spectrum at momentum regions <1.5\,GeV/c and >10\,GeV/c is larger than 15\,$\%$.  
The systematic errors on the $\cos\theta^{\prime}$ and $\phi^{\prime}$
distributions for $\mu^{-}$ and $\mu^{+}$ are shown in
Figs~\ref{fig:fluxthetasyst} (a) to (d). 
The obtained systematic errors due to different sources in the zenith
and azimuthal spectra are less than 5\,$\%$. 
\section{Results and Discussion}
\label{sec:results}
The weighted true distribution of $q/p$, $\cos\theta^{\prime}$ and
$\phi^{\prime}$ at the last iteration represents the generated level
distribution is a better representation of the
observed data.
The y-axis in the weighted true distributions is in
the unit of counts per area of the RPC for the total data taking
period. The weighted true distribution of momentum, $\cos\theta^{\prime}$ and $\phi^{\prime}$ at the
last iteration represents the generated level distribution
which can better reproduce the observed reconstructed data. This section elaborates the conversion of the weighted true
distribution of $\mu^{-}$ and $\mu^{+}$ from the last iteration to the
differential spectrum in proper normalised flux units as a function
of momentum, $\cos\theta^{\prime}$ and $\phi^{\prime}$.

\subsection{Estimation of Differential Muon Flux as a Function of Momentum}
\label{subsec:momflux}

The weighted true momentum distributions of $\mu^{-}$ and $\mu^{+}$ are
converted into differential fluxes by normalising to the area of the RPC, solid angle
acceptance, total live time, data acquisition efficiency,  and momentum bin
width. The differential flux as a function of momentum $p$ is defined as

\begin{equation}
\Phi^{\pm}(p) =
\frac{N^{\mathrm{weighted}}_{\mathrm{Gen}(\pm)}(p)}
{A \times \Omega \times T_{\mathrm{total}} \times
\epsilon_{\mathrm{DAQ}} \times \Delta p},
\label{eqn:pfluxnorm}
\end{equation}

where $N^{\mathrm{weighted}}_{\mathrm{Gen}(\pm)}(p)$ is the number of
weighted muons in a given momentum bin $p$, $A$ is the area of
the RPC detector, and $\Omega$ is the solid-angle factor defined as

\[
\Omega = 2\pi \int_{0^\circ}^{\theta^{\prime}_{f}}
\cos^{2}\theta^{\prime} \sin\theta^{\prime} \, d\theta^{\prime}.
\]

Here, $T_{\mathrm{total}}$ represents duration of the total data-taking period
(in seconds), $\epsilon_{\mathrm{DAQ}}$ is the data acquisition
efficiency, $\Delta p$ is the momentum bin width and $\theta^{\prime}_{f}$ is
the maximum zenith angle (11.48$^{\circ}$, 19.95$^{\circ}$
and 25.84$^{\circ}$). 
The resulting differential momentum spectra for $\mu^{-}$ and
$\mu^{+}$ for different zenith angle acceptances are shown in Figs.~\ref{fig:fluxmom}(a) to (f),
respectively and also the numerical values are given in the Appendix. The
systematic error band is shown in grey colour, and the total error,
including the statistical error, is shown as a red-coloured band. The
statistical uncertainties in the reconstructed data are propagated to the
generated level flux using the Singular Value Decomposition
method using TDecompSVD in CERN-ROOT\cite{cernroot}. The present measurements are compared with previous 
experimental results from CAPRICE94\cite{caprice},
CAPRICE97\cite{caprice}, and BESS-TeV\cite{bess-tev}. The observation
site details for CAPRICE94 are Lynn Lake,
Manitoba, Canada (56.5$^{\circ}$\,N, 101.0$^{\circ}$\,W), at an altitude of
360\,m above sea level, and a nominal vertical geomagnetic
cutoff of 0.5\,GV. The CAPRICE97 was performed at Fort Sumner, New
Mexico, (34.3$^{\circ}$\,N, 104.1$^{\circ}$\,W), at an altitude of
1270\,m and a vertical cutoff
of 4.2\,GV. The measured momentum spectrum from CAPRICE94 and
CAPRICE97 considers the maximum zenith angle of the reconstructed
muons to be 20$^{\circ}$. 
BESS-TeV ground observation was done at Tsukuba, Japan
(36.2$^{\circ}$\,N, 140.1$^{\circ}$\,W), 30\,m above sea level with a
vertical cutoff rigidity of 11.4\,GV. BESS-TeV analysis used the
reconstructed muon angle of $\cos\theta^{\prime}$>0.98 for the estimation of
momentum spectrum.  
Overall, the observed spectral shapes are consistent
with earlier observations. However, the measured flux in almost all
the energy regions is systematically lower than several previous
measurements. 
This suppression is primarily attributed to the relatively large
geomagnetic cutoff rigidity (17\,GV) at the present observation site,
which reduces the contribution from low- and intermediate-rigidity
primary cosmic rays responsible for secondary muon production.
In addition to the comparison with other experimental results, predictions from various
low-energy hadronic interaction models implemented in CORSIKA are
scaled using the primary cosmic ray fluxes.
The spectral index of cosmic muons in CORSIKA predictions
differs from the observation.

\begin{figure}[htbp]
\centering
\includegraphics[width=0.99\linewidth]{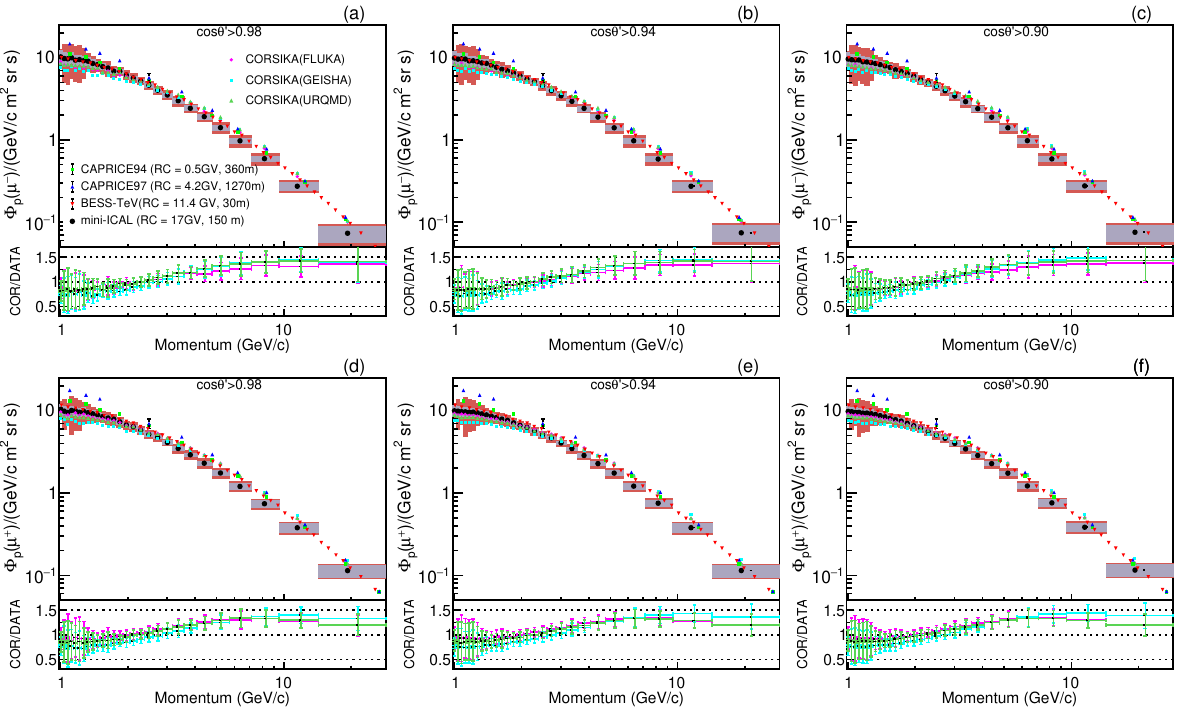}
\caption{\label{fig:fluxmom} Differential energy spectra of cosmic muons for (a-c) $\mu^{-}$ and (d-f) $\mu^{+}$ for the different acceptance of zenith angles.}
\end{figure}

\subsection{Estimation of Zenith and Azimuthal Spectra of Muons}

For completeness, the muon flux is also evaluated as a function of
$\cos\theta^{\prime}$ and $\phi^{\prime}$.
The vertical integrated fluxes for $\mu^{-}$ and $\mu^{+}$,
denoted by $I_{0}^{-}$ and $I_{0}^{+}$, are estimated using the near vertical
muons (0.99\,$\leq$\,$\cos\theta^{\prime}$\,$\leq$\,1.0 or
0\,$\leq$\,$\theta^{\prime}$\,$\leq$\,8.109$^{\circ}$) from the
weighted true $\cos\theta^{\prime}$ distributions. The total number of events
is normalised using the RPC detector area ($A$), solid-angle acceptance $\Omega$,
$T_{\mathrm{total}}$, and $\epsilon_{\mathrm{DAQ}}$.

\begin{figure}[htbp]
\centering
\includegraphics[width=0.99\linewidth]{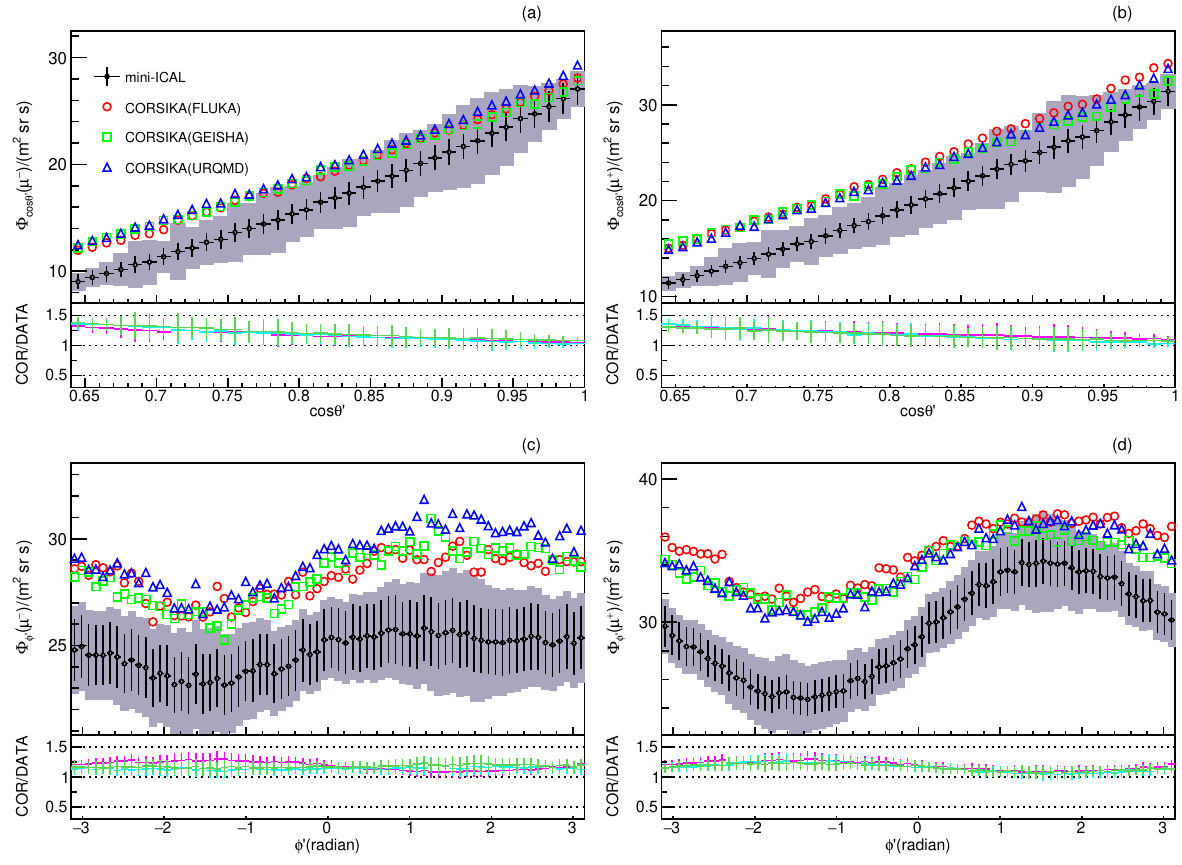}
\caption{\label{fig:fluxcosthephi} (a) and (b) are the zenith-angle spectra of cosmic muons for $\mu^{-}$ and $\mu^{+}$ respectively. (c) and (d) are the azimuth-angle spectra of cosmic muons for $\mu^{-}$ and $\mu^{+}$ respectively.}
\end{figure}

The resulting vertical fluxes are

\[
I_{0}^{-} = 27.06 \pm 1.47 \, \mathrm{m^{-2}\,sr^{-1}\,s^{-1}},
\quad
I_{0}^{+} = 31.39 \pm 1.48 \, \mathrm{m^{-2}\,sr^{-1}\,s^{-1}},
\]

yielding a combined vertical flux of

\[
I_{0} = 58.45 \pm 2.08 \, \mathrm{m^{-2}\,sr^{-1}\,s^{-1}}.
\]
The following vertical fluxes are estimated from the different low-energy models FLUKA, GHEISHA and URQMD in CORSIKA,
\[
I_{0}^{-} (FLUKA)  =  28.0899\, \mathrm{m^{-2}\,sr^{-1}\,s^{-1}},
\quad
I_{0}^{+} (FLUKA) =  34.2892\, \mathrm{m^{-2}\,sr^{-1}\,s^{-1}},
\]
\[
I_{0}^{-} (GHEISHA)  = 27.8328 \, \mathrm{m^{-2}\,sr^{-1}\,s^{-1}},
\quad
I_{0}^{+} (GHEISHA) =  32.4698 \, \mathrm{m^{-2}\,sr^{-1}\,s^{-1}},
\]
\[
I_{0}^{-} (URQMD)  =  29.3074 \, \mathrm{m^{-2}\,sr^{-1}\,s^{-1}},
\quad
I_{0}^{+} (URQMD) =  33.7866 \, \mathrm{m^{-2}\,sr^{-1}\,s^{-1}},
\]
The present value of total vertical flux is $\sim$\,6\,\%, $\sim$\,3\,\%
and $\sim$\,7\,\%, lower than the FLUKA, GHEISHA and URQMD
respectively. 
The normalised zenith-angle spectra with respect to their maximum counts of
$\mu^{-}$ and $\mu^{+}$, scaled using the respective vertical fluxes $I_{0}^{-}$ and
$I_{0}^{+}$, are shown in Figs.~\ref{fig:fluxcosthephi}(a) and (b)
respectively, where the systematic uncertainty is shown as an error bar
and the total uncertainty, including the statistical fluctuation, is shown
as a grey-colored band.  
The shape of the zenith angle flux in the data is falling faster than the
CORSIKA predictions, which might be a result of uncertainties in the
atmospheric modelling and the spectral index of the primary cosmic ray
momentum spectrum. 

The azimuthal distributions are normalised using the detector area,
solid-angle factor, total live time, and DAQ efficiency. The effective
solid angle for each azimuthal bin is defined as

\[
\Omega_{\phi^{\prime}} =
\frac{2\pi}{N_{\phi^{\prime}}}
\int_{0^\circ}^{50^\circ}
\cos^{2}\theta^{\prime} \sin\theta^{\prime} \, d\theta^{\prime},
\]

where $N_{\phi^{\prime}}$ is the number of azimuthal bins.
The resulting azimuthal fluxes for $\mu^{-}$ and $\mu^{+}$, along with
predictions from CORSIKA simulations, are presented in
Fig.~\ref{fig:fluxcosthephi} (c) and (d), respectively. The overall shape of the $\mu^{-}$
distribution is broadly consistent with model expectations.
However, the observed azimuthal asymmetry at the experimental site,
particularly the East--West effect for $\mu^{+}$, is more pronounced
than predicted by the nominal CORSIKA simulations.
The lower vertical muon flux and the enhanced east--west asymmetry
observed in the data compared with the CORSIKA predictions may be
related to uncertainties in the geomagnetic cutoff rigidity used in
the simulation, as well as deviations of the primary cosmic-ray
spectrum from a single power-law description and uncertainties in its
chemical composition, as indicated by measurements from AMS-02\cite{ams2015_1,ams2015_2,ams2021_3}, CALET\cite{calet2022,calet2023},
and DAMPE\cite{dampe2019, dampe2021,dampe2026}.
A dedicated study incorporating these effects is required
to quantify their contributions.


\section{Conclusions}
\label{sec:concl}
We have presented a detailed measurement of the charge-separated
cosmic muon flux near the geomagnetic equator using the prototype
mini-ICAL detector operated at the IICHEP-Transit Campus, Madurai (altitude 150\,m;
latitude 9.9372$^\circ$\,N; longitude 78.013$^\circ$\,E; geomagnetic
latitude 1.44$^\circ$\,N, vertical geomagnetic cutoff rigidity of 17\,GV).
The analysis is based on data collected between 12 December 2018 and
28 December 2018 under both non-magnetic and magnetic configurations.
Given the finite momentum resolution of mini-ICAL up to 5\,GeV/c, an
iterative forward-folding approach was adopted in place of
conventional unfolding. The generated-level spectra were
parameterised and iteratively tuned until convergence was achieved.
Stable solutions were obtained within a few iterations, demonstrating
the robustness of the forward-folding framework in extracting true
distributions in detectors with moderate resolution.
The differential fluxes of $\mu^{-}$ and $\mu^{+}$ were estimated in
the energy range $\sim$\,1--28\,GeV/c. The spectral shapes are broadly
consistent with previous measurements such as CAPRICE and BESS-TeV.
However, the flux is systematically lower than that
reported at sites with smaller geomagnetic cutoff rigidities. This
suppression is naturally explained by the large (17\,GV) cutoff at
the present site, which reduces the contribution of low- and
intermediate-rigidity primary cosmic rays responsible for secondary
muon production.
Comparisons with different low-energy hadronic interaction models
implemented in CORSIKA reveal noticeable discrepancies in the momentum
spectrum.
These differences highlight the importance of improved modelling of
hadronic interactions and geomagnetic effects at low geomagnetic
latitudes.
Overall, this work demonstrates the capability of the mini-ICAL
detector to measure charge-dependent atmospheric muon spectra with
controlled detector systematics. The forward-folding methodology
developed here provides a reliable framework for extracting
generated-level distributions in experiments with limited momentum
resolution.
Therefore, this measurement offers valuable
input for improving atmospheric muon and neutrino flux calculations at
low geomagnetic latitudes and for testing hadronic interaction models
in the few-GeV and higher-energy regime, which is directly relevant to
future precision neutrino oscillation experiments. 
Although the systematic uncertainties in the estimated flux are larger
in low momentum, the central value in this momentum region is also
sensitive to uncertainties in the modelling of the atmospheric density
profile and the geomagnetic field. Thus the observed excess of the measured
flux over the model predictions at low momentum should therefore be
interpreted in the context of these combined atmospheric and
geomagnetic effects, rather than being attributed solely to
differences in hadronic interaction models.

\appendix

\acknowledgments
We gratefully acknowledge the invaluable support of the India-based
Neutrino Observatory (INO). We sincerely thank all the members of the
INO project, the Inter-Institutional Centre for
High Energy Physics (IICHEP) Transit campus, Madurai, the Tata Institute of
Fundamental Research (TIFR), Mumbai, and the Bhabha Atomic Research
Centre (BARC), Mumbai, whose dedicated efforts were instrumental in
the construction, commissioning, and operation of the mini-ICAL
detector. We also acknowledge Dr. P. K. Mohanty and Dr. B. Hariharan, GRAPES lab, TIFR, for providing the rigidity
cutoff map for the observation site. 



\section{Appendix}
\begin{table}[h]
\centering
\begin{tabular}{|c|c|c|c|c|c|}
\hline
Bin No & $\Delta p$ & $p_{avg}$ &
$\Phi_{\mu^{-}}$  ($\cos\theta^{\prime}$>0.98) & $\Phi_{\mu^{-}}$
($\cos\theta^{\prime}$>0.94) & $\Phi_{\mu^{-}}$  ($\cos\theta^{\prime}$>0.90)\\
\hline
 & (GeV/c) & (GeV/c) &
1/(GeV/c m$^{2}$ sr s) &
1/(GeV/c m$^{2}$ sr s) &
1/(GeV/c m$^{2}$ sr s) \\
\hline
  1  & 1.02--1.06 & 1.04 & 9.75\,$\pm$\,4.75 & 9.59\,$\pm$\,4.66 & 9.42\,$\pm$\,4.58 \\ \hline
  2  & 1.06--1.1 & 1.08 & 9.66\,$\pm$\,4.96 & 9.67\,$\pm$\,4.94 & 9.37\,$\pm$\,4.8 \\ \hline
  3  & 1.1--1.14 & 1.12 & 9.96\,$\pm$\,4.34 & 9.61\,$\pm$\,4.19 & 9.36\,$\pm$\,4.08 \\ \hline
  4  & 1.14--1.19 & 1.17 & 9.3\,$\pm$\,4.23 & 9.13\,$\pm$\,4.16 & 8.90\,$\pm$\,4.05 \\ \hline
  5 & 1.19--1.24 & 1.22 & 9.46\,$\pm$\,4.52 & 9.16\,$\pm$\,4.38 & 8.85\,$\pm$\,4.23 \\ \hline
  6  & 1.24--1.3 & 1.27 & 9.18\,$\pm$\,3.58 & 8.95\,$\pm$\,3.49 & 8.69\,$\pm$\,3.39 \\ \hline
  7  & 1.3--1.36 & 1.33 & 9.04\,$\pm$\,2.28 & 8.61\,$\pm$\,2.16 & 8.48\,$\pm$\,2.14 \\ \hline
  8  & 1.36--1.43 & 1.39 & 8.49\,$\pm$\,1.86 & 8.51\,$\pm$\,1.86 & 8.22\,$\pm$\,1.8 \\ \hline
  9  & 1.43--1.5 & 1.47 & 8.24\,$\pm$\,1.35 & 8.07\,$\pm$\,1.33 & 7.90\,$\pm$\,1.31 \\ \hline
  10  & 1.5--1.59 & 1.54 & 7.78\,$\pm$\,1.27 & 7.65\,$\pm$\,1.25 & 7.60\,$\pm$\,1.24 \\ \hline
  11 & 1.59--1.68 & 1.63 & 7.49\,$\pm$\,1.55 & 7.22\,$\pm$\,1.50 & 7.12\,$\pm$\,1.48 \\ \hline
  12 & 1.68--1.79 & 1.73 & 6.85\,$\pm$\,1.04 & 6.90\,$\pm$\,1.00 & 6.79\,$\pm$\,1.03 \\ \hline
  13 & 1.79--1.9 & 1.84 & 6.79\,$\pm$\,0.72 & 6.52\,$\pm$\,0.70 & 6.40\,$\pm$\,0.70 \\ \hline
  14 & 1.9--2.04 & 1.97 & 6.14\,$\pm$\,0.76 & 6.01\,$\pm$\,0.74 & 5.95\,$\pm$\,0.74 \\ \hline
  15 & 2.04--2.2 & 2.12 & 5.64\,$\pm$\,0.69 & 5.47\,$\pm$\,0.68 & 5.41\,$\pm$\,0.67 \\ \hline
  16 & 2.2--2.38 & 2.29 & 5.20\,$\pm$\,0.53 & 5.10\,$\pm$\,0.52 & 4.97\,$\pm$\,0.51 \\ \hline
  17 & 2.38--2.6 & 2.48 & 4.62\,$\pm$\,0.55 & 4.56\,$\pm$\,0.55 & 4.49\,$\pm$\,0.54 \\ \hline
  18 & 2.6--2.86 & 2.72 & 4.02\,$\pm$\,0.48 & 3.98\,$\pm$\,0.48 & 3.96\,$\pm$\,0.47 \\ \hline
  19 & 2.86--3.17 & 3.01 & 3.51\,$\pm$\,0.43 & 3.44\,$\pm$\,0.42 & 3.41\,$\pm$\,0.42 \\ \hline
  20 & 3.17--3.57 & 3.36 & 2.99\,$\pm$\,0.42 & 2.94\,$\pm$\,0.41 & 2.92\,$\pm$\,0.41 \\ \hline
  21 & 3.57--4.08 & 3.81 & 2.42\,$\pm$\,0.26 & 2.42\,$\pm$\,0.26 & 2.40\,$\pm$\,0.25 \\ \hline
  22 & 4.08--4.76 & 4.40 & 1.92\,$\pm$\,0.26 & 1.89\,$\pm$\,0.25 & 1.89\,$\pm$\,0.25 \\ \hline
  23 & 4.76--5.71 & 5.20 & 1.41\,$\pm$\,0.20 & 1.40\,$\pm$\,0.20 & 1.41\,$\pm$\,0.20 \\ \hline
  24 & 5.71--7.14 & 6.36 & 0.975\,$\pm$\,0.171 & 0.976\,$\pm$\,0.171 & 0.979\,$\pm$\,0.171 \\ \hline
  25 & 7.14--9.52 & 8.18 & 0.592\,$\pm$\,0.095 & 0.586\,$\pm$\,0.094 & 0.588\,$\pm$\,0.094 \\ \hline
  26 & 9.52--14.3 & 11.50 & 0.274\,$\pm$\,0.046 & 0.274\,$\pm$\,0.045 & 0.276\,$\pm$\,0.046 \\ \hline
  27 & 14.3--28.6 & 19.30 & 0.0735\,$\pm$\,0.0210 & 0.0744\,$\pm$\,0.0213 & 0.0756\,$\pm$\,0.0217 \\ \hline
\end{tabular}
\caption{The measured muon flux at different momentum bins for $\mu^{-}$. The quoted error represents the includes both statistical and systematics uncertainties. }
\label{tab:sample}
\end{table}

\begin{table}[h]
\centering
\begin{tabular}{|c|c|c|c|c|c|}
\hline
Bin No & $\Delta p$ & $p_{avg}$ &
$\Phi_{\mu^{+}}$ ($\cos\theta^{\prime}$>0.98) & $\Phi_{\mu^{+}}$
($\cos\theta^{\prime}$>0.94) & $\Phi_{\mu^{+}}$  ($\cos\theta^{\prime}$>0.90) \\
\hline
 & (GeV/c) & (GeV/c) &
1/(GeV/c m$^{2}$ sr s) &
1/(GeV/c m$^{2}$ sr s) &
1/(GeV/c m$^{2}$ sr s) \\
\hline
   1 & 1.02--1.06 & 1.04 & 9.86\,$\pm$\,3.88 & 9.86\,$\pm$\,3.87 & 9.72\,$\pm$\,3.82 \\ \hline
   2 & 1.06--1.1 & 1.08 & 9.51\,$\pm$\,2.84 & 9.73\,$\pm$\,2.90 & 9.53\,$\pm$\,2.84 \\ \hline
   3 & 1.1--1.14 & 1.12 & 10.07\,$\pm$\,4.96 & 9.67\,$\pm$\,4.78 & 9.54\,$\pm$\,4.72 \\ \hline
   4 & 1.14--1.19 & 1.17 & 9.97\,$\pm$\,4.18 & 9.57\,$\pm$\,4.03 & 9.35\,$\pm$\,3.95 \\ \hline
   5 & 1.19--1.24 & 1.22 & 9.56\,$\pm$\,3.78 & 9.6\,$\pm$\,3.8 & 9.35\,$\pm$\,3.69 \\ \hline
   6 & 1.24--1.3 & 1.27 & 9.80\,$\pm$\,2.82 & 9.43\,$\pm$\,2.72 & 9.21\,$\pm$\,2.66 \\ \hline
   7 & 1.3--1.36 & 1.33 & 9.46\,$\pm$\,2.39 & 9.14\,$\pm$\,2.32 & 9.00\,$\pm$\,2.30 \\ \hline
   8 & 1.36--1.43 & 1.39 & 9.16\,$\pm$\,1.91 & 8.89\,$\pm$\,1.86 & 8.70\,$\pm$\,1.83 \\ \hline
   9 & 1.43--1.5 & 1.47 & 8.89\,$\pm$\,1.52 & 8.61\,$\pm$\,1.48 & 8.43\,$\pm$\,1.46 \\ \hline
   10 & 1.5--1.59 & 1.54 & 8.48\,$\pm$\,1.25 & 8.26\,$\pm$\,1.21 & 8.11\,$\pm$\,1.20 \\ \hline
   11 & 1.59--1.68 & 1.63 & 8.16\,$\pm$\,1.31 & 7.91\,$\pm$\,1.29 & 7.78\,$\pm$\,1.27 \\ \hline
   12 & 1.68--1.79 & 1.73 & 7.89\,$\pm$\,1.19 & 7.64\,$\pm$\,1.16 & 7.49\,$\pm$\,1.15 \\ \hline
   13 & 1.79--1.9 & 1.84 & 7.31\,$\pm$\,1.03 & 7.13\,$\pm$\,1.0 & 7.02\,$\pm$\,0.99 \\ \hline
   14 & 1.9--2.04 & 1.97 & 6.71\,$\pm$\,0.76 & 6.63\,$\pm$\,0.77 & 6.56\,$\pm$\,0.77 \\ \hline
   15 & 2.04--2.2 & 2.12 & 6.38\,$\pm$\,0.74 & 6.25\,$\pm$\,0.73 & 6.13\,$\pm$\,0.72 \\ \hline
   16 & 2.2--2.38 & 2.29 & 5.78\,$\pm$\,0.62 & 5.65\,$\pm$\,0.61 & 5.58\,$\pm$\,0.60 \\ \hline
   17 & 2.38--2.6 & 2.48 & 5.16\,$\pm$\,0.54 & 5.09\,$\pm$\,0.53 & 5.06\,$\pm$\,0.53 \\ \hline
   18 & 2.6--2.86 & 2.72 & 4.76\,$\pm$\,0.55 & 4.67\,$\pm$\,0.53 & 4.61\,$\pm$\,0.53 \\ \hline
   19 & 2.86--3.17 & 3.01 & 4.11\,$\pm$\,0.37 & 4.05\,$\pm$\,0.36 & 4.00\,$\pm$\,0.35 \\ \hline
   20 & 3.17--3.57 & 3.36 & 3.47\,$\pm$\,0.41 & 3.47\,$\pm$\,0.40 & 3.42\,$\pm$\,0.39 \\ \hline
   21 & 3.57--4.08 & 3.81 & 2.91\,$\pm$\,0.34 & 2.86\,$\pm$\,0.33 & 2.85\,$\pm$\,0.33 \\ \hline
   22 & 4.08--4.76 & 4.40 & 2.30\,$\pm$\,0.30 & 2.28\,$\pm$\,0.29 & 2.27\,$\pm$\,0.29 \\ \hline
   23 & 4.76--5.71 & 5.20 & 1.75\,$\pm$\,0.24 & 1.74\,$\pm$\,0.24 & 1.74\,$\pm$\,0.24 \\ \hline
   24 & 5.71--7.14 & 6.36 & 1.20\,$\pm$\,0.17 & 1.21\,$\pm$\,0.17 & 1.22\,$\pm$\,0.16 \\ \hline
   25 & 7.14--9.52 & 8.18 & 0.740\,$\pm$\,0.101 & 0.752\,$\pm$\,0.102 & 0.758\,$\pm$\,0.101 \\ \hline
   26 & 9.52--14.3 & 11.50 & 0.378\,$\pm$\,0.063 & 0.379\,$\pm$\,0.063 & 0.384\,$\pm$\,0.063 \\ \hline
   27 & 14.3--28.6 & 19.30 & 0.114\,$\pm$\,0.022 & 0.114\,$\pm$\,0.022 & 0.116\,$\pm$\,0.022 \\ \hline
\end{tabular}
\caption{The measured muon flux at different momentum bins for $\mu^{+}$. The quoted error represents the includes both statistical and systematics uncertainties.}
\label{tab:sample2}
\end{table}



\end{document}